\documentclass[aps,prl,amssymb,amsmath,nobibnotes,nofootinbib,superscriptaddress,notitlepage,showkeys,twocolumn]{revtex4-2}

\usepackage{graphicx}
\usepackage{bm}
\usepackage{color}
\usepackage{xspace}
\usepackage{ulem}
\newcommand{\fermi}{\emph{Fermi}\xspace}

\newcommand{\lat}{\emph{Fermi}-LAT\xspace}

\newcommand{\gr}{$\gamma$-ray\xspace}
\newcommand{\grs}{$\gamma$ rays\xspace}

\renewcommand{\deg}{\ensuremath{^{\circ}}\xspace}

\renewcommand\apj{Astrophys J}
\newcommand\apjs{Astrophys J Suppl Ser}

\newcommand\mnras{Mon Not R Astron Soc}
\newcommand\jcap{J Cosmol Astropart Phys}

\newcommand\apjl{Astrophys J Lett}

\newcommand\raa{Res Astron Astrophys}

\begin{document}

\title{Search for gamma-ray spectral lines with the DArk Matter Particle Explorer}

\author{Francesca~Alemanno}
\affiliation{Gran Sasso Science Institute (GSSI), I-67100 L'Aquila, Italy}
\affiliation{Istituto Nazionale di Fisica Nucleare (INFN) - Laboratori Nazionali del Gran Sasso, I-67100 L'Aquila, Italy}

\author{Qi~An}
\affiliation{State Key Laboratory of Particle Detection and Electronics, University of Science and Technology of China, Hefei 230026, China}
\affiliation{Department of Modern Physics, University of Science and Technology of China, Hefei 230026, China}

\author{Philipp~Azzarello}
\affiliation{Department of Nuclear and Particle Physics, University of Geneva, CH-1211 Geneva, Switzerland}

\author{Felicia~Carla~Tiziana~Barbato}
\affiliation{Gran Sasso Science Institute (GSSI), I-67100 L'Aquila, Italy}
\affiliation{Istituto Nazionale di Fisica Nucleare (INFN) - Laboratori Nazionali del Gran Sasso, I-67100 L'Aquila, Italy}

\author{Paolo~Bernardini}
\affiliation{Dipartimento di Matematica e Fisica E. De Giorgi, Universit\`a del Salento, I-73100 Lecce, Italy}
\affiliation{Istituto Nazionale di Fisica Nucleare (INFN) - Sezione di Lecce, I-73100 Lecce, Italy}

\author{Xiao-Jun~Bi}
\affiliation{Particle Astrophysics Division, Institute of High Energy Physics, Chinese Academy of Sciences, Beijing 100049, China}
\affiliation{University of Chinese Academy of Sciences, Beijing 100049, China}

\author{Ming-Sheng~Cai}
\affiliation{Key Laboratory of Dark Matter and Space Astronomy, Purple Mountain Observatory, Chinese Academy of Sciences, Nanjing 210023, China}
\affiliation{School of Astronomy and Space Science, University of Science and Technology of China, Hefei 230026, China}

\author{Elisabetta~Casilli}
\affiliation{Dipartimento di Matematica e Fisica E. De Giorgi, Universit\`a del Salento, I-73100 Lecce, Italy}
\affiliation{Istituto Nazionale di Fisica Nucleare (INFN) - Sezione di Lecce, I-73100 Lecce, Italy}

\author{Enrico~Catanzani}
\affiliation{Istituto Nazionale di Fisica Nucleare (INFN) - Sezione di Perugia, I-06123 Perugia, Italy}

\author{Jin~Chang}
\affiliation{Key Laboratory of Dark Matter and Space Astronomy, Purple Mountain Observatory, Chinese Academy of Sciences, Nanjing 210023, China}
\affiliation{School of Astronomy and Space Science, University of Science and Technology of China, Hefei 230026, China}

% ...

\collaboration{DAMPE Collaboration}
\email{dampe@pmo.ac.cn}
\noaffiliation

\author{Yun-Feng~Liang}
\affiliation{Laboratory for Relativistic Astrophysics, Department of Physics, Guangxi University, Nanning 530004, China}

\begin{abstract}
   The DArk Matter Particle Explorer (DAMPE) is well suitable for searching for monochromatic and sharp $\gamma$-ray structures in the GeV$-$TeV range thanks to its unprecedented high energy resolution.
    In this work, we search for $\gamma$-ray line structures using five years of DAMPE data.
    To improve the sensitivity, we develop two types of dedicated data sets (including the BgoOnly data which is the first time to be used in the data analysis for the calorimeter-based gamma-ray observatories) and adopt the signal-to-noise ratio optimized regions of interest (ROIs) for different DM density profiles.
    No line signals or candidates are found between 10 and 300 GeV in the Galaxy.
The constraints on the velocity-averaged cross section for $\chi\chi \to \gamma\gamma$ and the decay lifetime for $\chi \to \gamma\nu$, both at 95\% confidence level, have been calculated and the systematic uncertainties have been taken into account.
    Comparing to the previous \lat results, though  DAMPE has an acceptance smaller by a factor of $\sim 10$, similar constraints on the DM parameters are achieved and  below 100~GeV the lower limits on the decay lifetime are even stronger by a factor of a few.
    Our results demonstrate the potential of  high-energy-resolution observations on dark matter detection.
\end{abstract}


\maketitle

\section{Introduction}\label{sec:introduction}
The modeling of the cosmic microwave background fluctuations suggests that
in the Universe the average energy density of matter is roughly 5.4 times larger
than the baryon density~\citep{Aghanim2018}.
The discrepancy is usually explained with an extra cold matter component---the dark matter (DM).
DM is also required to explain other phenomena observed at different scales, such as the rotation curve of galaxies, the large mass-to-luminosity ratio of galaxy clusters, and the spatial offset of the center of the total mass from the center of the baryonic mass in the Bullet Cluster (see Refs. \citep{Bergstrom2000,Bertone2018,Clowe:2006eq} and references therein).
So far, however, the nature of the DM is still unclear.
The weakly interacting massive particles (WIMPs) are a leading candidate for cold DM, since they provide a natural explanation for the observed DM relic density~\citep{Jungman1996,Feng2010}.
If two WIMPs ($\chi$) can annihilate into a photon ($\gamma$) and another particle ($X$) directly, an approximately monochromatic structure at $E_\gamma = m_\chi\,(1-m_X^2/4m_\chi^2)$ will be produced (in the case of decay one should replace $m_\chi$ with $m_\chi/2$).
Such processes have been proposed in some extensions of the standard model of particle physics,
such as the lightest supersymmetric particles annihilating through $\chi\chi \to \gamma\gamma$ or $\chi\chi\to\gamma Z^0$~\citep{Bergstrom1988,Rudaz1989,Ullio1998},
or the gravitinos decaying through $\chi \to \gamma\nu$ with $R$ parity violation~\citep{Ibarra2008}.
Besides, peak-like spectral features may also arise from the virtual internal bremsstrahlung process in the DM annihilation~\citep{Beacom2005,Bergstrom:2004cy} or the decay of low-mass intermediate particles generated by the annihilating/decaying DM~\citep{Ibarra2012}.
Since these distinct spectral features are hard to be produced in known astrophysical processes, a robust detection would be a smoking-gun signature of WIMPs.

Though plentiful works have been carried out to search for \gr lines after the launch of \fermi Gamma-ray Space Telescope, no GeV line signal has been formally detected yet~\citep{FermiLAT2009, Abdo2010,Vertongen2011,Ackermann2012,Bringmann2012,Weniger2012,Tempel2012,Su2012a,Su2012,Huang2012,Hektor2013,Albert2014,Ackermann2015,Anderson2016,Liang2016,Liang2016b,Liang2017,LiS2019,Mazziotta2020,Shen2021}.
The potential $\sim 133~\rm GeV$ line reported in the Galactic center~\citep{Bringmann2012,Weniger2012,Su2012a} was recognized as a systematic effect later~\citep{Ackermann2015}.
The tentative signature at $\sim 43~\rm GeV$ from a nearby Galaxy cluster sample~\citep{Liang2016} is found to have a strange temporal behavior of its significance \citep{Shen2021}.
Nevertheless, the \gr line signal is so important that independent line searches with data from different telescopes are needed.

The DArk Matter Particle Explorer (DAMPE) is a space-borne high energy particle detector launched on 17 December 2015.
It aims to measure charged cosmic rays and \grs in a very wide energy range~\citep{Chang2014,Chang2017,Ambrosi2019}.
From the top to bottom, DAMPE consists of a Plastic Scintillator strip Detector (PSD), a Silicon-Tungsten tracKer-converter (STK), a Bismuth Germanium Oxide (BGO) imaging calorimeter and a NeUtron Detector (NUD).
The PSD measures the particle charge and acts as an anti-coincidence detector.
The STK converts the incident \gr photons to electron pairs and records the trajectories.
The BGO measures the energies of incident particles and images the profile of shower.
The NUD further enhances the electron/proton separation.

The BGO calorimeter has a thickness of 32 radiation length, with which the deposit energy of electron/\gr events can be effectively absorbed and the shower developments can be well contained. As a result, for electrons/\grs, the energy resolution of DAMPE  is significantly higher than \lat in a wide energy range~\citep{Chang2017}.
Since a better energy resolution will not only make the line structure more evident in the spectrum, but also reduce the systematic uncertainties, DAMPE is well suitable for searching for the monochromatic spectral structures.
In this work, we perform a line search using the DAMPE \gr observations of the inner Galaxy, set constraints on the DM parameters, and demonstrate the  potential of high-energy-resolution observations on dark matter detection.

\section{Data Selection}\label{sec:data}
The local significance of a line-like structure can be approximated by $n_{\rm line}/\sqrt{n_{\rm bkg,eff}}$~\citep{Ackermann2013a,Chernoff1954},
where $n_{\rm line}$ (the photon counts of a line) is proportional to the acceptance $\mathcal{A}$, and
$n_{\rm bkg,eff}$ is the number of background events below the line, i.e., the effective background reported in previous works~\citep{Ackermann2013a,Albert2014,Ackermann2015}.
Approximately we have $n_{\rm bkg,eff} \sim N \times \int \min\{f_{\rm bkg}(E'), f_{\rm sig}(E'; E_{\rm line})\}\,{\rm d} E'$, where $N$ is the total background counts in the fit range, $f_{\rm bkg}$ and $f_{\rm sig}$ are the background and signal probability density functions, respectively~\citep{Ackermann2013a}.
For a very narrow energy dispersion profile, $n_{\rm bkg,eff}$ reduces to $N\,f_{\rm bkg}(E_{\rm line})\,\Delta E$, which is proportional to the acceptance $\mathcal{A}$ and the energy resolution $\Delta E/E$.
Therefore, the significance of a line improves when the division of acceptance and energy resolution $\sqrt{\mathcal{A}/(\Delta E/E)}$ increases.

In our analysis, two dedicated data sets, namely the LineSearch and BgoOnly data sets~\citep{Xu2021}, are developed and combined to improve the sensitivity.
The former contains the events converted in the STK.
Compared with the standard STK converting events based on the algorithm in Ref.~\citep{Xu2018}, these events are required to pass through more BGO layers to improve the sensitivity of lines by maximizing $\mathcal{A}/(\Delta E/E)$.
The latter contains the events converted in the BGO calorimeter, in which photons are reconstructed based on the tracks in the BGO detector.
The incident energies of all the events are reconstructed from the deposit energies in the calorimeter using a parameterized correction method~\citep{Yue2017}.
The total acceptance of these data sets is $\sim 1600~\rm cm^2\,sr$ at 5~GeV and $\sim 1900~\rm cm^2\,sr$ between 10 and 100~GeV.
The 68\% containment of energy resolution averaged over the acceptance is smaller than 1.7\% (1.0\%) above 10~GeV (35~GeV) for both types of data~\citep{Xu2021}. To our knowledge, this is the first time to use the BgoOnly data in the analysis for the calorimeter-based \gr observatories.

In this work, the above data sets from Jan. 1, 2016 to Dec. 31, 2020 are chosen.
The energy range is restricted from 5 to 450~GeV with the DAMPE \gr science toolkit {\tt DmpST}~\citep{Duan2019,Jiang2020}.
We only choose the events satisfying the High-Energy Trigger (HET) condition.
Data collected during the South Atlantic Anomaly or strong solar flares has been excluded.
In total, over 90 thousand \gr events are used for the analysis.

\begin{figure}
    \centering
    \includegraphics[width=0.48\textwidth]{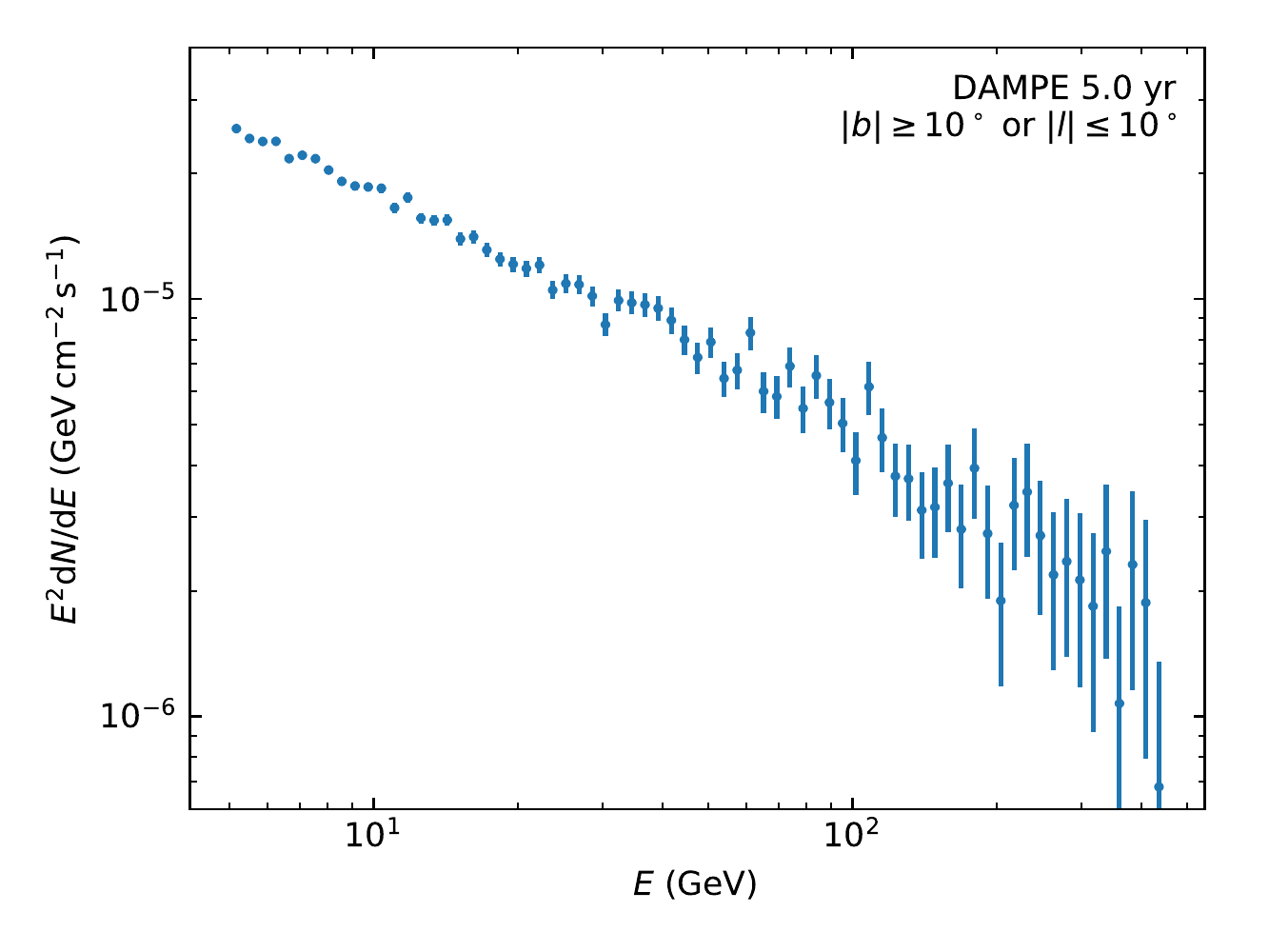}
    \caption{\label{fig:flux}
        (Color online)
        The average SED of the LineSearch and BgoOnly photons from the region with Galactic plane removed.
    }
\end{figure}

Based on the live time during the observation and the Monte Carlo (MC) instrument response functions, we are able to calculate the exposure and energy dispersion profiles.
Fig.~\ref{fig:flux} shows the average spectral energy distributions (SEDs) of the region with the Galactic plane ($|l|>10^\circ$ and $|b|<10^\circ$, where $l$ and $b$ are longitude and latitude in the Galactic coordinate, respectively)~\citep{Abdo2010} removed.
The SED is almost featureless and no obvious line-like structure displays.
To be quantitative, we perform an unbinned analysis in the following.

\section{Methodology}\label{sec:analysis}
DM density profile $\rho_{\rm DM}$ is uncertain particularly in the inner Galaxy, so we consider three representative profiles, including
the Navarro-Frenk-White (NFW) profile $\rho_{\rm NFW}(r)=\rho_s/[(r/r_s)(1+r/r_s)^2]$ with $r_s=20~\rm kpc$~\citep{Navarro1996}, the Einasto profile $\rho_{\rm Ein}(r)=\rho_s\exp\{-(2/\alpha)[(r/r_s)^\alpha-1]\}$ with $r_s=20~\rm kpc$ and $\alpha=0.17$~\citep{Einasto1965,Navarro2010}, and the isothermal profile $\rho_{\rm iso}(r) = \rho_s/[1+(r/r_s)^2]$ with $r_s=5~\rm kpc$~\citep{Bahcall1980}.
The normalization $\rho_s$ is governed by  $\rho_{\rm DM}(R_0)=0.4~\rm GeV\,cm^{-3}$~\citep{Catena2010} and $R_0=8.5~\rm kpc$~\citep{Ghez2008}.

For both the annihilation and decay scenarios, we make the regions of interest (ROIs) optimized for the sensitivity, where we approximate the recorded photon counts as the spatial distribution of the background, and multiply the exposure with different DM density profiles for the anticipated signal.
All of the ROIs are circular regions with radius $R_{\rm GC}$ centering at the Galactic center but with the rectangular region $|l|\geq\Delta l$ and $|b|\leq5\deg$ masked.
For the annihilating DM, the optimal $(R_{\rm GC}, \Delta l)$ are $(16\deg,~ 5\deg)$, $(40\deg, ~9\deg)$ and $(86\deg,~ 0\deg)$ for the Einasto, NFW and isothermal profiles, respectively.
For the decaying DM, all the optimal $(R_{\rm GC}, \Delta l)$ for different profiles are close to $(150\deg, ~ 0\deg)$, so this parameter set is adopted as a representative.
Table~\ref{tab:ROIs} presents the ROIs for different density profiles and the corresponding $J$-factors $J_{\rm DM}=\int_{\rm ROI} {\rm d}\Omega \int {\rm d}s\,\rho_{\rm DM}^2$ (annihilating DM) or $D$-factors $D_{\rm DM}=\int_{\rm ROI} {\rm d}\Omega \int {\rm d}s\,\rho_{\rm DM}$ (decaying DM), where $s$ is the distance along the line of sight.

\begin{table}
    \centering
    \caption{
        The signal-to-noise optimal ROIs for three different DM density profiles for annihilating DM or decaying DM. The corresponding $J$-factors for annihilating DM and $D$-factors for decaying DM are also presented, whose units are $\rm GeV^2\,cm^{-5}$ and $\rm GeV\,cm^{-2}$ respectively.
    }\label{tab:ROIs}
    \begin{tabular}{lcccc}
        \hline
                &\multicolumn{2}{c}{Annihilation}&\multicolumn{2}{c}{Decay} \\
        Profile & $\rm ROI$ & $J$-factor & $\rm ROI$ & $D$-factor \\
        \hline
        Einasto    & R16  & $9.00\times 10^{22}$ & R150 & $2.42\times 10^{23}$ \\
        NFW        & R40  & $9.50\times 10^{22}$ & R150 & $2.40\times 10^{23}$ \\
        Isothermal & R86  & $6.58\times 10^{22}$ & R150 & $2.53\times 10^{23}$ \\
        \hline
    \end{tabular}
\end{table}

We perform an unbinned likelihood analysis with the sliding-window technique to quantify the significance of the hypothesized lines, which will mitigate the bias caused by the background spectral shape and energy binning.
For a line at $E_{\rm line}$, only the photons in the window from $0.5\,E_{\rm line}$ to $1.5\,E_{\rm line}$ are used in the fittings.
The energy difference between two adjacent windows is $0.5\,\sigma_E$, where $\sigma_E$ is the half width of the 68\% exposure weighted energy dispersion containment in the Galactic center for LineSearch data set~\citep{Ackermann2015,Liang2016}.
To make sure the Chernoff's theorem valid~\citep{Chernoff1954}, at least 30 photons are required in each window, which restricts the highest line energy in the ROI R16 to 211~GeV.
The unbinned likelihood function $L_k(\Theta)$ for the data set $k$ in the energy window of $[E_{\rm min}, E_{\rm max}]$ is defined as
\begin{equation}
    \ln L_k(\Theta)=\sum_{i=1}^{n_{k}} \ln [ \bar{\lambda}_k(E_i; \Theta) ] - {\int_{E_{\rm min}}^{{E_{\rm max}}}} \bar{\lambda}_k(E; \Theta)\,{\rm d}E,
\end{equation}
where $n_{k}$ is the number of observed photons in the given energy range, and $\bar{\lambda}_k(E; \Theta)$ is the expected counts in model per energy range with the parameter $\Theta$, which is calculated with the exposure $\bar{\epsilon}_k(E)$ at energy $E$ averaged over the ROI.
The likelihood to be fitted is $L(\Theta)=L_1(\Theta)\times L_2(\Theta)$, where the subscript indices represent two data sets.

In each energy window, a likelihood ratio test~\cite{Cash1979} is performed.
The null hypothesis consists of a power-law background, i.e., $\bar{\lambda}_{{\rm null},k}(E; \Theta) = F_{\rm b}(E)\,\bar{\epsilon}_k(E)$, while the signal hypothesis contains a monochromatic line and a power-law background, i.e., $\bar{\lambda}_{{\rm sig},k}(E) = F_{\rm b}(E)\,\bar{\epsilon}_k(E) + \bar{F}_{{\rm s},k}(E)\,\bar{\epsilon}_k(E_{\rm line})$.
The power-law spectrum and the line structure are defined as $F_{\rm b}(E;N_{\rm b},\Gamma)=N_{\rm b}\,E^{-\Gamma}$ and $\bar{F}_{{\rm s},k}(E;N_{\rm s},E_{\rm line})=N_{\rm s}\,\bar{D}_{{\rm eff},k}(E;E_{\rm line})$ (i.e., $S_{\rm line}(E)=N_{\rm s}\,\delta(E-E_{\rm line})$ before convolution), respectively, where $N_{\rm s}$ is non-negative.
$\bar{D}_{\rm eff}$ is the exposure weighted energy dispersion function averaged over the ROI and is given by
\begin{equation}
    \bar{D}_{{\rm eff},k}(E; E_{\rm line})=\frac{\sum_{ij}D_k(E; E_{\rm line}, \theta_j)\,\epsilon_k(E_{\rm line}, \theta_j, {\bf r}_i)}{\sum_{ij} \epsilon_k(E_{\rm line}, \theta_j, {\bf r}_i)},
\end{equation}
where $\theta$ is the incident angle with respect to the boresight, $\bf r$ is the pixel coordinate within the ROI, and $D(E)$ is the energy dispersion function of DAMPE~\citep{Duan2019}.
We fit both models to the data using the {\tt MINUIT}~\cite{MINUIT1975} and then calculate the test statistic (TS) value ${\rm TS} \equiv -2\,\ln (\hat{L}_{\rm null} / \hat{L}_{\rm sig})$, where $\hat{L}_{\rm null}$ and $\hat{L}_{\rm sig}$ are the maximum likelihood values of the null and alternative model respectively.

\section{Results}\label{sec:results}
\begin{figure}
    \centering
    \includegraphics[width=0.48\textwidth]{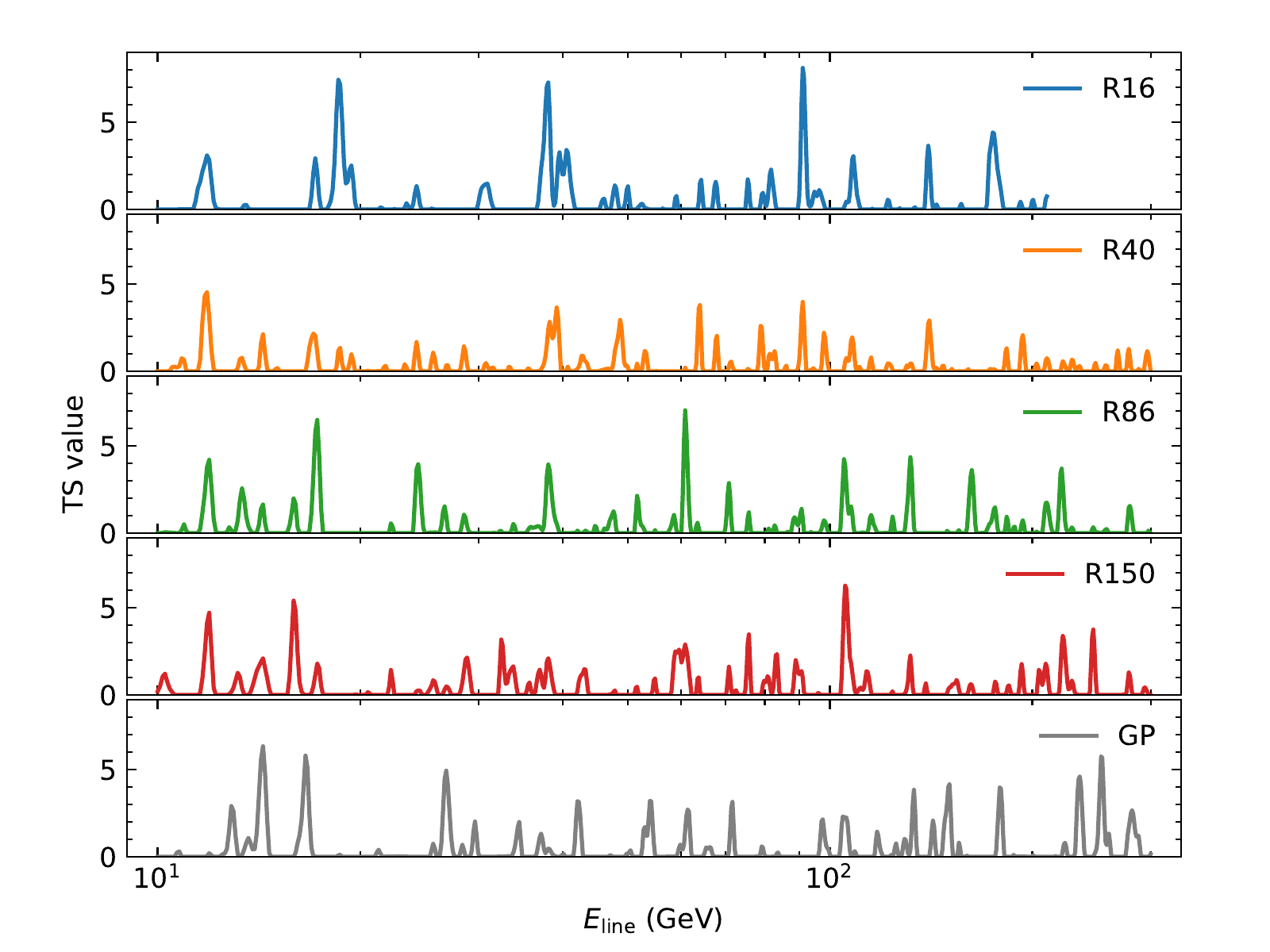}
    \caption{\label{fig:ts}
        (Color online)
        TS values of line candidates at various energies in the signal optimized ROIs and the Galactic plane region ($|l|>30\deg$ and $|b|<5\deg$).
        The local significance can be calculated with $s_{\rm local}=\sqrt{\rm TS}$~\citep{Ackermann2013a}.
        The highest line energy for R16 ROI is 211~GeV which is limited by the minimum photon counts required in a window.
    }
\end{figure}

We do not find any \gr line signal or candidate (${\rm TS} \geq 9$) in all the ROIs (Fig.~\ref{fig:ts}).
Therefore we calculate the 95\% confidence level (C.L.) constraints on the DM parameter space.

\begin{figure*}
    \centering
    \includegraphics[width=0.48\textwidth]{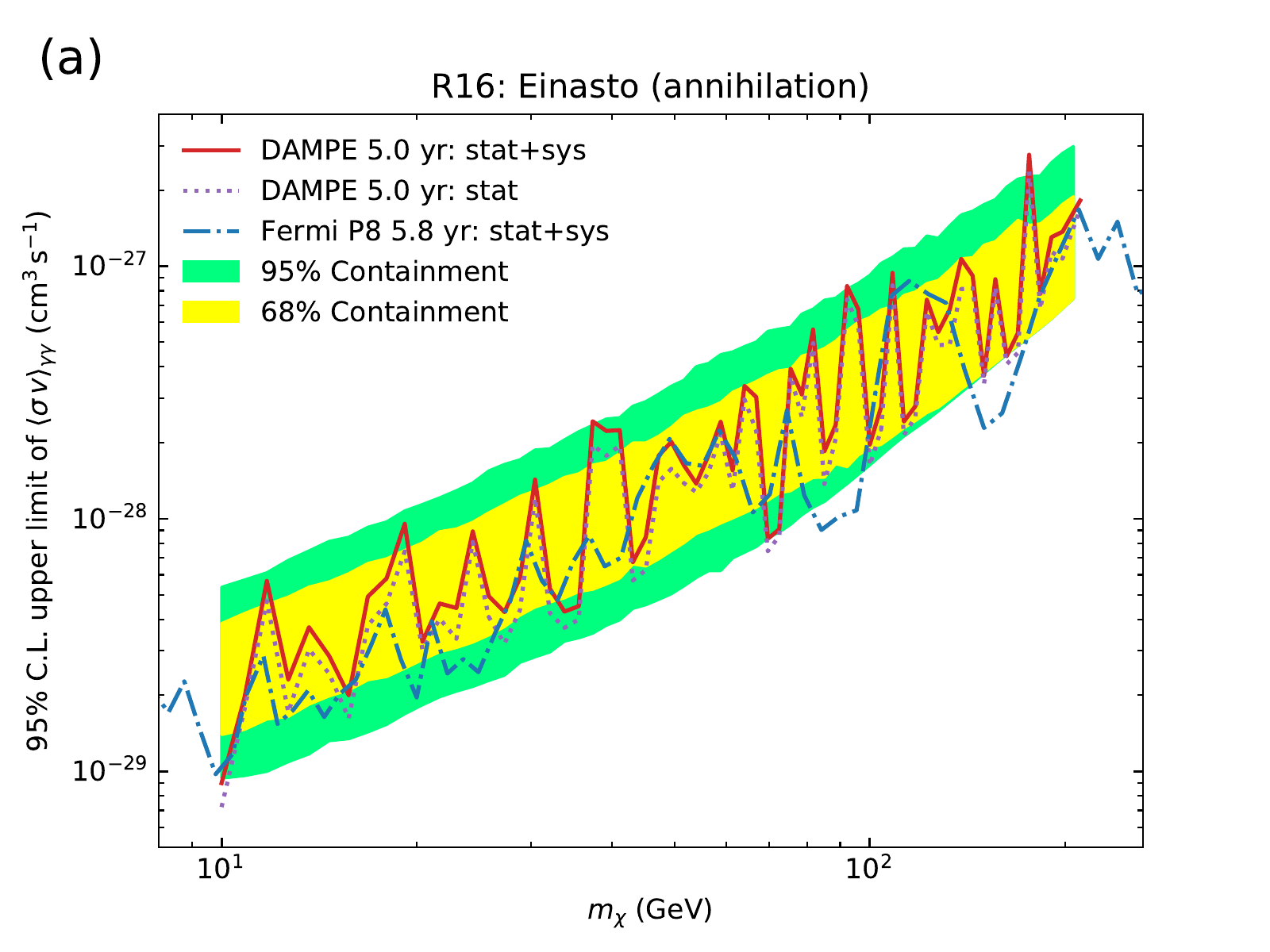}
    \includegraphics[width=0.48\textwidth]{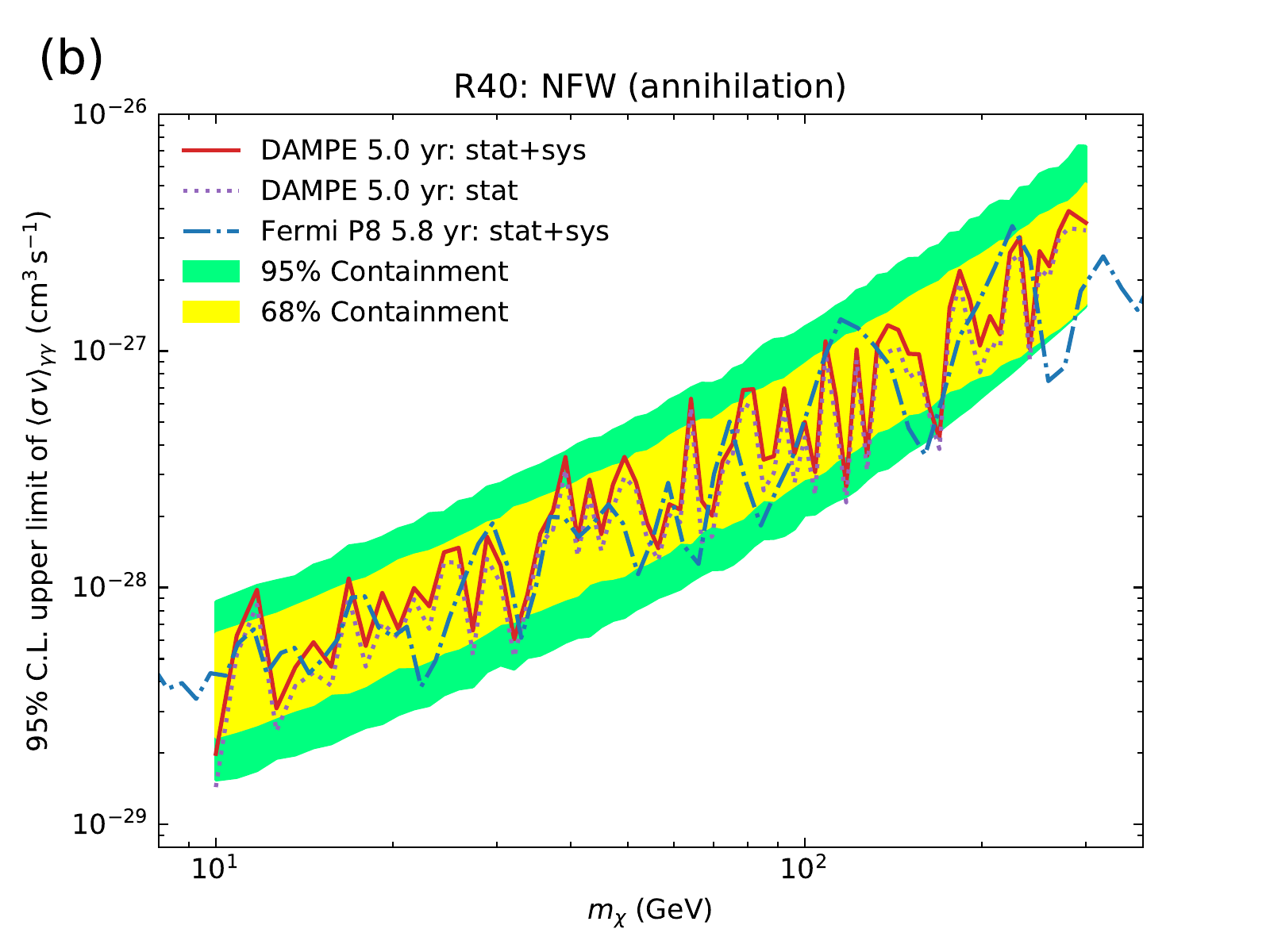}
    \includegraphics[width=0.48\textwidth]{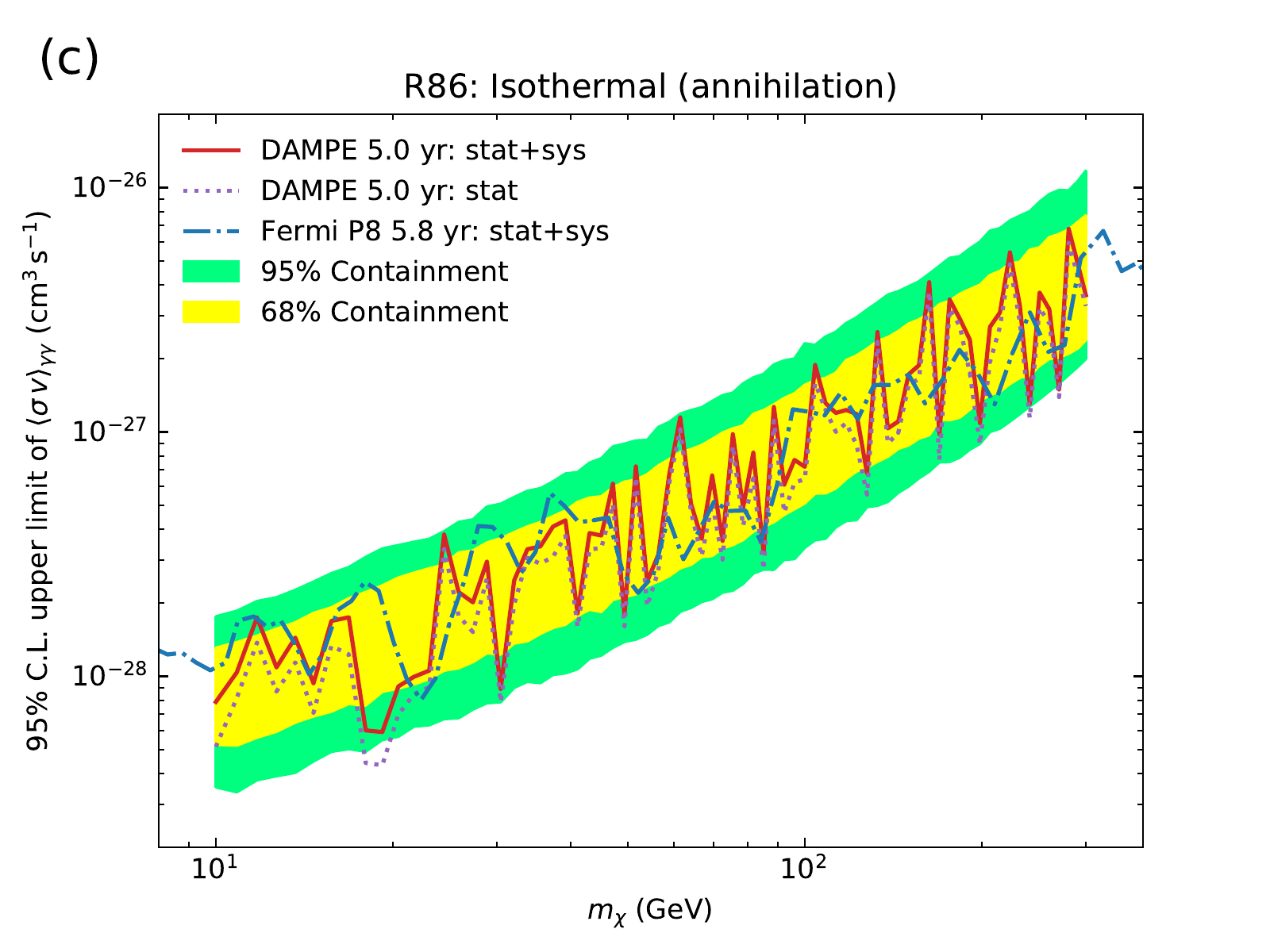}
    \includegraphics[width=0.48\textwidth]{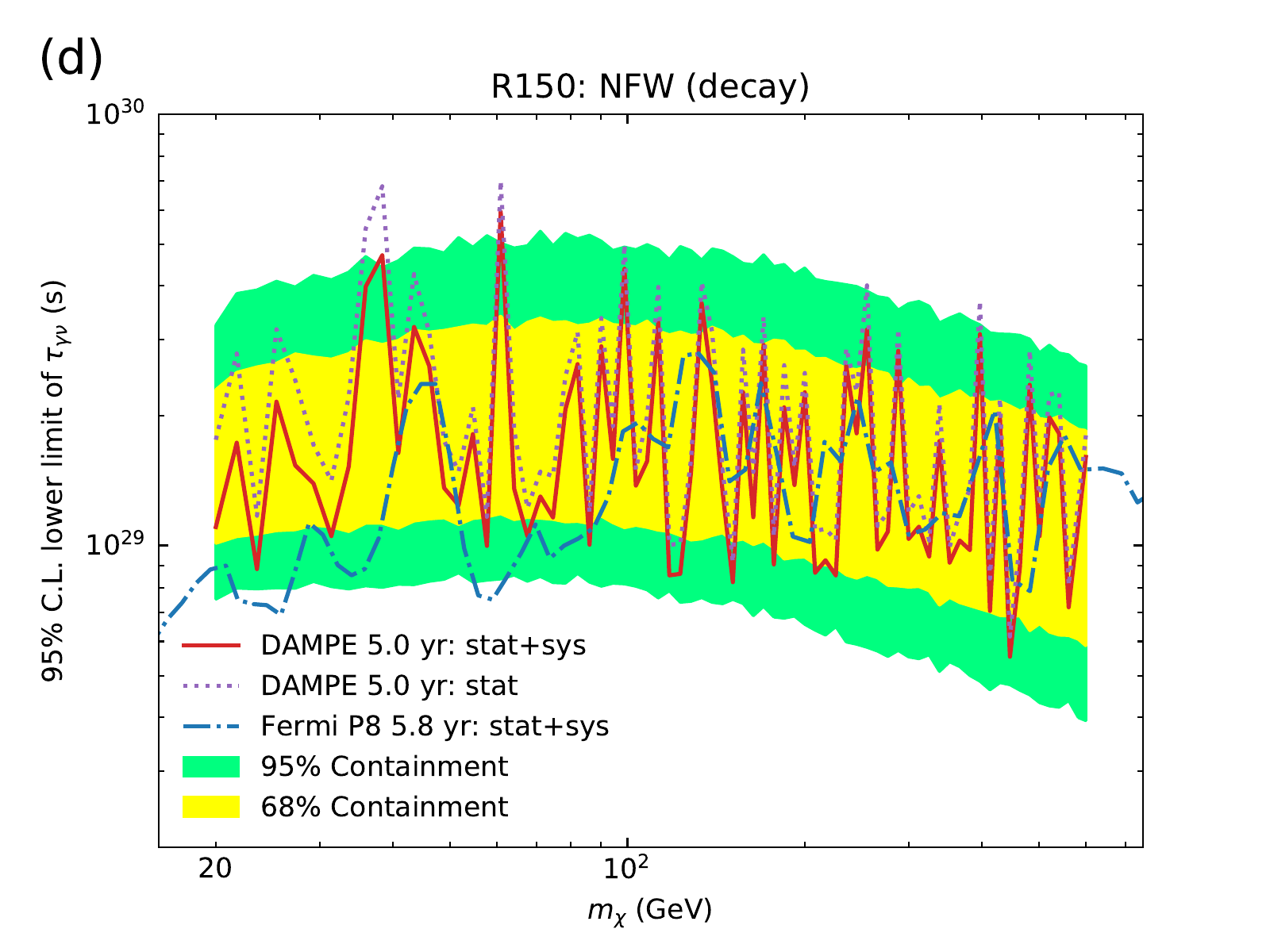}
    \caption{\label{fig:sv_tau}
        The 95\% C.L. constraints for different DM density profiles.
        (a--c) The $\left< \sigma v\right>_{\gamma\gamma}$ upper limits of annihilating DM assuming the (a) Einasto, (b) NFW and (c) Isothermal profile respectively.
        (d) The $\tau_{\gamma\nu}$ lower limit of decaying DM assuming the NFW profile.
        Yellow (green) bands show the 68\% (95\%) expected containment obtained from 1000 simulations of background emission with systematic uncertainties involved.
        The red solid and purple dotted lines are the results with and without the systematic uncertainties respectively.
        The blue dot-dashed lines show the 5.8-year \lat constraints~\citep{Ackermann2015}.
    }
\end{figure*}

For annihilating DM, the \gr spectrum is given by
\begin{equation}
    S_{\rm line}(E) = \frac{1}{4\pi} \frac{J_{\rm DM}\times\left< \sigma v\right>_{\gamma\gamma}}{2m_\chi^2}\,2\delta(E-E_{\rm line}),
\end{equation}
where $\left< \sigma v\right>_{\gamma\gamma}$ is the velocity-averaged annihilation cross section for $\chi\chi \to \gamma\gamma$, and $m_\chi$ is the rest mass of a DM particle which satisfies $E_{\rm line}=m_\chi$.
For the decaying DM, the spectrum can be written as
\begin{equation}
    S_{\rm line}(E) = \frac{1}{4\pi} \frac{D_{\rm DM}}{m_\chi\,\tau_{\gamma\nu}}\,\delta(E-E_{\rm line}),
\end{equation}
where $\tau_{\gamma\nu}$ is the lifetime of a DM particle through the $\chi \to \gamma\nu$ process and $E_{\rm line}=0.5 m_\chi$.
We increase (decrease) the cross section (lifetime) from its best-fit value (under the condition of $N_{\rm s} \ge 0$) until the log-likelihood changes by 1.35 to achieve the constraints.
These results are conservative since we do not take into account the contributions from extragalactic DM annihilation or decay~\citep{Bergstrom:2001jj,Blanco:2018esa}.

Purple dotted lines in the Fig.~\ref{fig:sv_tau} show the $\left< \sigma v\right>_{\gamma\gamma}$ and $\tau_{\gamma\nu}$ constraints for various DM density profiles without the systematic uncertainties.
The cuspy profiles show better constraints since they have larger $J$-factors and better background-to-noise ratios.
Because of smaller energy dispersion and low statistics of DAMPE data, more fluctuations appear in the constraints compared with the previous results.

Three types of systematic uncertainties are considered in this work~(see also Ref. \citep{Ackermann2013a}):

(1) Uncertainties about the conversion from the signal counts to the fluxes, which are mainly associated with the uncertainties of exposure.
The overall uncertainty of effective area is $|\delta\epsilon/\epsilon| \lesssim 2.5\%$ between 2 and 10~GeV based on the ratio of event fraction in the flight data to the MC prediction of Vela pulsar~\citep{Ackermann2012_Calibration}.
The $|\delta\epsilon/\epsilon|$ caused by the spatial binning of exposure is less than 2\% and 5\% for R16 and R150 respectively, if compared to the results calculated with smaller pixel bins of 0.5\deg.
Adding them quadratically, we have $|\delta\epsilon/\epsilon| \lesssim 6\%$.

(2) Uncertainties that could affect the expected signal counts.
The uncertainty $\delta n_{\rm sig}/n_{\rm sig}$ arised from the width of sliding windows is on average 1.5\% if the narrow boundaries of $0.7\,E_{\rm line}$ -- $1.3\,E_{\rm line}$ are used.
The uncertainty from the shared background emission parameters in likelihood functions of the two data sets is on average $< 0.3\%$.
The systematic uncertainty of energy resolution is $\lesssim 20\%$ if we compare the MC results to the beam test ones~\citep{Chang2017} which will thereby lead to the mean signal counts uncertainty of $8\%$--$9\%$.
The overall systematic uncertainties of signal counts $\delta n_{\rm sig}/n_{\rm sig}$ are calculated at all the line energies, and the average value is approximately 9\%.

(3) Uncertainties that could mask or produce line-like structures.
The fractional signal $f \equiv n_{\rm sig}/b_{\rm eff}$ is often used to evaluate this type of uncertainties, where $n_{\rm sig}$ and $b_{\rm eff}$ are the signal counts and effective background counts, respectively~\citep{Ackermann2013a,Ackermann2015}.
The analysis of the control regions along the Galactic plane between 10 and 14~GeV shows $|\delta f_{\rm sys}| = 1.3\%$ (refer to the Supplementary materials for detail).
The fraction of cosmic ray contamination is $\lesssim 1.5\%$ above 10~GeV, as found in the simulations~\citep{Xu2018}.
Therefore the overall systematic part of fractional signal is $|\delta f_{\rm sys}| \lesssim 2.0\%$.

To incorporate the systematic uncertainties, we replace the likelihood function with~\citep{Albert2014,Buckley2015}
\begin{equation}
    L(n_{\rm sig}) \to L(n_{\rm sig}+n_{\rm sys})\times \frac{1}{\sqrt{2\pi}\sigma_{\rm sys}} \exp\left(-\frac{n_{\rm sys}^2}{2 \sigma_{\rm sys}^2}\right),
    \label{eqn::like_sys}
\end{equation}
where $\sigma_{\rm sys}=|\delta f_{\rm sys}| \times b_{\rm eff}$ is the the standard deviation of systematic uncertainty, and $n_{\rm sys}$ describes the counts from the false signal.
The exposure and signal counts are also scaled according to the uncertainties.

Our 95\% C.L. constraints for different DM density profiles, after addressing the systematic uncertainties,  are shown in red solid lines in Fig.~\ref{fig:sv_tau}.
The expected 68\% and 95\% containment regions obtained with 1000 simulations of the best-fit power-law null model are also drawn in yellow and green bands, respectively, which encompass almost all the  fluctuations of the upper limits.
Most of our results are comparable to the 5.8-year results of \lat in blue dot-dashed lines with the systematic uncertainties included.
For the decaying DM, our lower limits on the decay lifetime are stronger than that from \lat by a factor of $\sim 2$ for DM with mass $\lesssim 100~\rm GeV$.
Even though DAMPE has a much smaller data set than \lat (DAMPE just has an  acceptance peaking at $\sim 0.2~{\rm m^2~sr}$~\citep{Xu2021}, which is smaller than that of \lat by a factor of $\sim 10$), a comparable or even better constraints are achieved for DAMPE because of the much higher energy resolution and the smaller impact of the systematic uncertainties, some of which are contributed by the components below the lines.

\section{Summary}\label{sec:summary}
DAMPE has an unprecedented high energy resolution due to its thick BGO calorimeter, and therefore has an advantage in detecting sharp structures.
In this work, we use 5.0 years of DAMPE data to search for spectral lines from 10 to 300 GeV.
To improve the sensitivity to line signals, two types of \gr data sets, the LineSearch and BgoOnly data sets, are developed. To our knowledge, this is the first time to take the BgoOnly data into the scientific analysis for the calorimeter-based \gr observatories (Previously, the production of calorimeter-only data was suggested by the \lat collaboration but so far such a kind of data is still unavailable).
We also make four ROIs optimized for DM density profiles for signals originating from the DM annihilation or decay in the Galaxy.
We use the summed unbinned likelihood function to combine two data sets and the sliding windows technique to reduce the uncertainty from the spectral shape of background emission.

No line signals or candidates with TS value $\geq 9$ are detected with 5.0 years of DAMPE data (Fig.~\ref{fig:ts}).
The 95\% C.L. constraints on the annihilation cross section or decay lifetime, with proper addressing of the systematic uncertainties, are presented.
As depicted in Fig.~\ref{fig:sv_tau},  most of our constraints are comparable to the 5.8-year results of \lat thanks to the better energy resolution and the smaller influence of the systematic uncertainties.
For the decaying DM, our lower limits on the decay lifetime are stronger for DM with mass $\lesssim 100~\rm GeV$ by a factor of $\sim 2$. In view of the fact that the DAMPE data set is about ten times smaller than that of {\it Fermi}-LAT, our findings demonstrate the potential of  high-energy-resolution observations on dark matter detection.

\vspace{3mm}
{\bf Conflict of interest:} The authors declare that they have no conflict of interest.

\vspace{3mm}
\begin{acknowledgments}
    {\bf Acknowledgments:}
    The DAMPE mission is funded by the Strategic Priority Science and Technology Projects in Space Science of Chinese Academy of Sciences.
    In China, the data analysis was supported in part by
    the National Key Research and Development Program of China (2016YFA0400200),
    the National Natural Science Foundation of China (U1738210, U1738206, 11921003, 12003074, U1738205, U1738207, U1738208, 12022302, 11773086, 12003069, 11903084, 11622327, and U1738123),
    the Scientific Instrument Developing Project of the Chinese Academy of Sciences (GJJSTD20210009),
    the Key Research Program of the Chinese Academy of Sciences (ZDRW-KT-2019-5),
    the Youth Innovation Promotion Association CAS,
    the Natural Science Foundation of Jiangsu Province (BK20201107),
    and the Entrepreneurship and Innovation Program of Jiangsu Province.
    In Europe, the activities and data analysis are supported by the Swiss National Science Foundation (SNSF), Switzerland, and the National Institute for Nuclear Physics (INFN), Italy.
\end{acknowledgments}

\vspace{3mm}
{\bf Author contributions:} This work is the result of the contributions and efforts of all the participating institutes. All authors have reviewed, discussed, and commented on the results and on the manuscript.
In line with the collaboration policy, the authors are listed alphabetically.

\bibliography{mybibs}

\vspace{3mm}
{\bf The DAMPE experiment:} The DAMPE is the first Chinese astronomical satellite, which consists of four sub-detectors, including the plastic scintillator detector, the silicon tracker, the BGO calorimeter and the neutron detector. As a general-purpose high-energy cosmic ray and gamma-ray detector, DAMPE is distinguished by the unprecedented high energy resolution on measuring the cosmic ray electrons and gamma-rays. The main scientific objectives addressed by DAMPE include probing the dark matter via the detection of high-energy electrons/positrons and gamma-rays, understanding the origin, acceleration and propagation of cosmic rays in the Milky Way, and studying the gamma-ray astronomy. The DAMPE mission is funded by the Strategic Priority Science and Technology Projects in Space Science of the Chinese Academy of Sciences. The DAMPE Collaboration consists of more than 140 members from 3 countries, including physicists, astrophysicists and engineers.

\clearpage

\setcounter{figure}{0}
\renewcommand\thefigure{S\arabic{figure}}
\setcounter{table}{0}
\renewcommand\thetable{S\arabic{table}}
\setcounter{equation}{0}
\renewcommand{\theequation}{S\arabic{equation}}

\section{Supplemental Material}




\subsection{\gr data sets}
Since photons only make up a small fraction of cosmic rays (CRs), an efficient algorithm is required to select photons from background events.
A sophisticated selection algorithm is developed for DAMPE data in~\citep{Xu2018}.

\begin{figure}[!htb]
    \centering
    \includegraphics[width=0.48\textwidth]{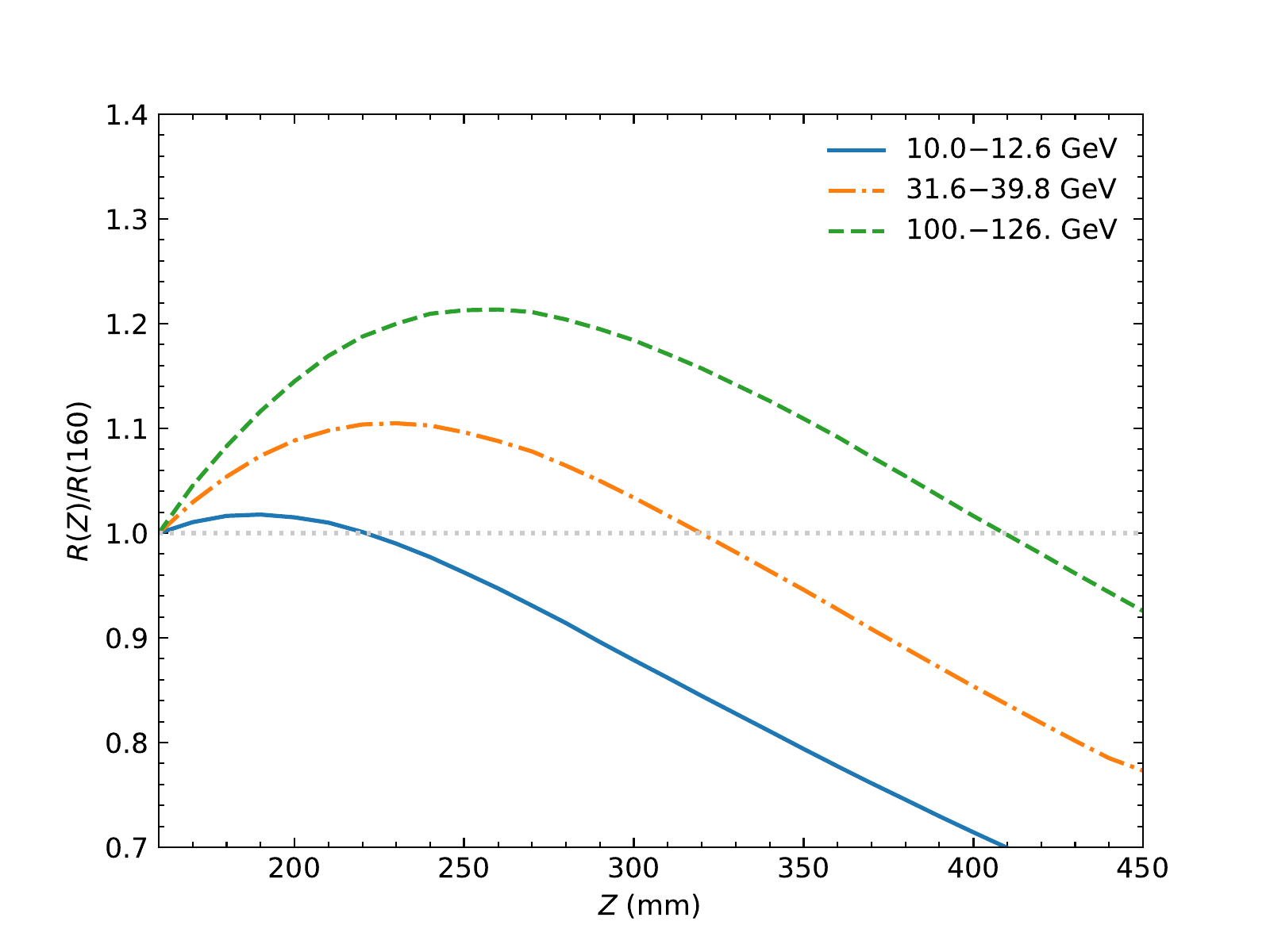}
    \caption{\label{fig:R_Z}
        The ratio of the $R$ of various $Z$s to $R$ of $Z=160~\mathrm{mm}$, where $R$ is the ratio of the acceptance to the energy resolution, and $Z$ is the minimum depth in event selection.
        The depth $Z=160~\rm mm$ is adopted in the standard data set~\citep{Xu2018}.
    }
\end{figure}

The standard data set aims to achieve a large photon acceptance and a strong background suppression level, so the track is required to pass through at least the first four layers of BGO calorimeter, which corresponds to the $Z$ direction value being more than $160~\rm mm$.
Since the sensitivity of linelike signal is related positively to the ratio of the acceptance to the energy resolution averaged over the acceptance $R\equiv\mathcal{A}/(\Delta E/E)$, we can make a subset of the standard data by maximizing the ratio $R$.
We first achieve the relation between $R$ and $Z$ by setting various minimum depths $Z$ to the MC data (Fig.~\ref{fig:R_Z}).
Optimal depths are found to be different for photons in different energies.
For low energy photons, because energies mostly deposit in the top few layers of BGO~\citep{Yue2017}, large $Z$ values will not improve the energy resolution too much but will always decrease the acceptance, thereby leading to small $R$.
On the other hand, particles from high energy photons are more likely to leak from four sides of the calorimeter, causing a poor energy resolution.
Requiring more BGO layers will improve the energy resolution significantly and may lead to better sensitivity.
To maximize the ratio $R$, we fit the optimal $Z$ values at different energies using a polynomial function $Z_{\rm poly}(E)$.
Only those STK events with $Z$ values larger than $Z_{\rm poly}(E)$ are selected to construct the LineSearch data set.

Approximately 40\% incident photons are converted into $e^+e^-$ pairs in the BGO calorimeter, therefore including these \gr events will improve the sensitivity.
Firstly, the same electron/proton separation as the standard data set is performed.
Since these events do not have STK tracks, we can only use the BGO tracks to provide direction information.
We require the events to pass through at least 8 BGO layers in order to achieve a track direction with a reasonable uncertainty.
The BGO track is extrapolated to the PSD considering the directional uncertainty and the charge rejection is made if sufficient energy is deposit in the corresponding PSD strips.
Applying the algorithm above to the BGO converting events, the BgoOnly data set is made.

More details can be found in \citep{Xu2021}.

\subsection{Instrument response functions (IRFs)}
\begin{figure*}[!htb]
    \centering
    \includegraphics[width=0.48\textwidth]{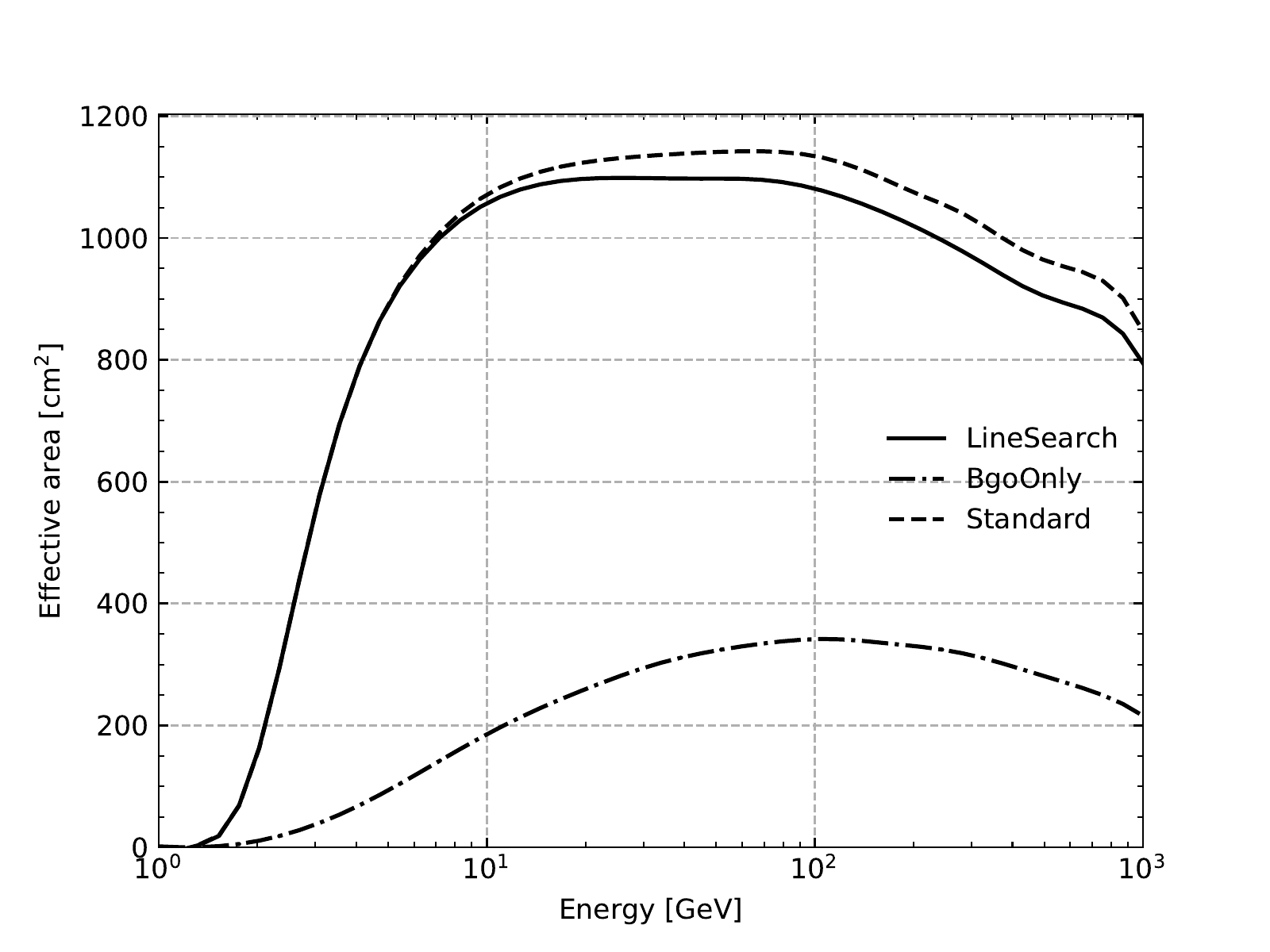}
    \includegraphics[width=0.48\textwidth]{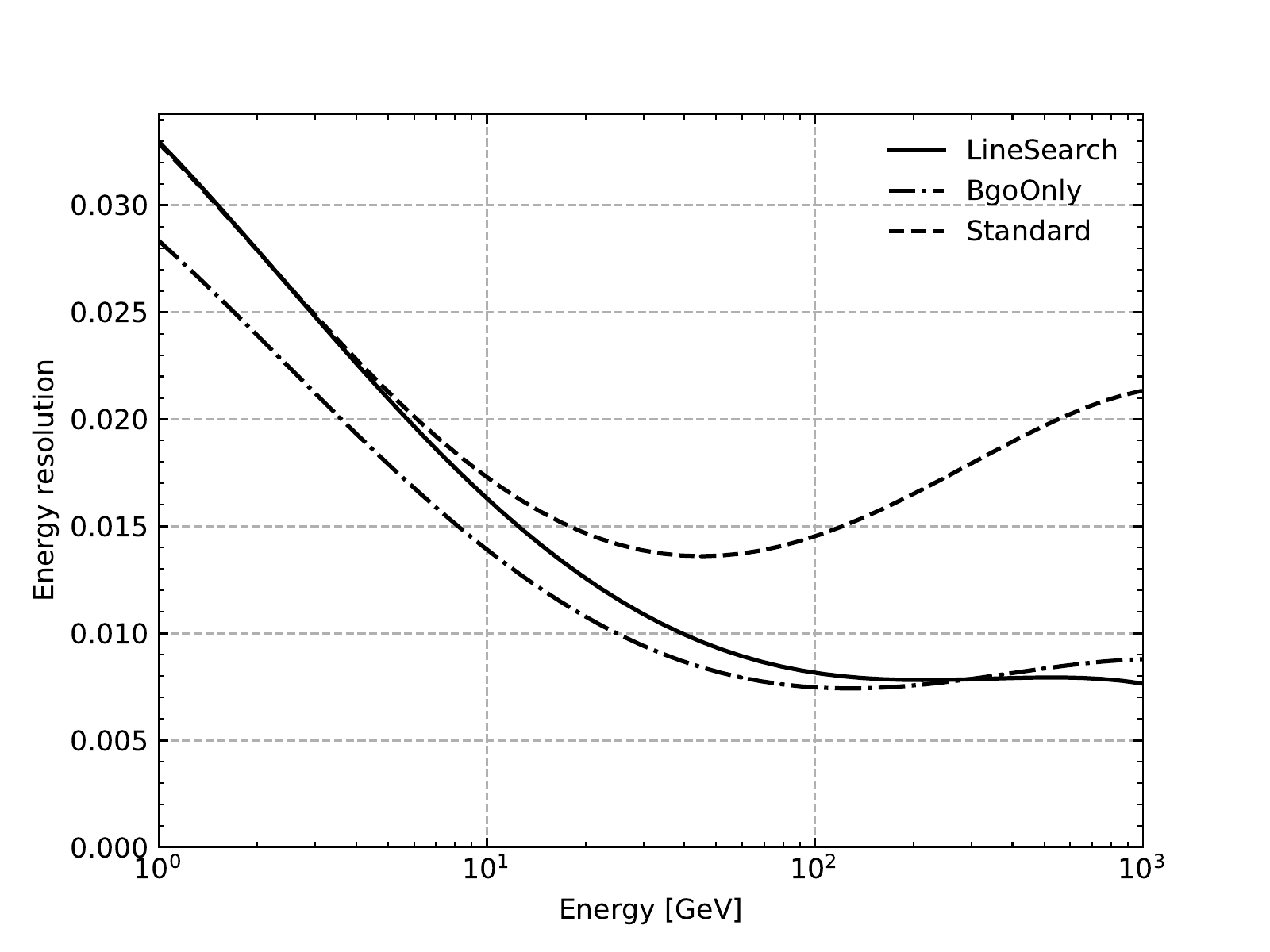}
    \caption{\label{fig:Aeff_edisp}
        The effective area at normal incidence (left) and energy resolution at $30\deg$ off-axis angle (right) for the LineSearch (solid), BgoOnly (dot-dashed) and Standard (dashed) data satisfying the HET condition.
    }
\end{figure*}

\begin{figure*}[!htb]
    \centering
    \includegraphics[width=0.48\textwidth]{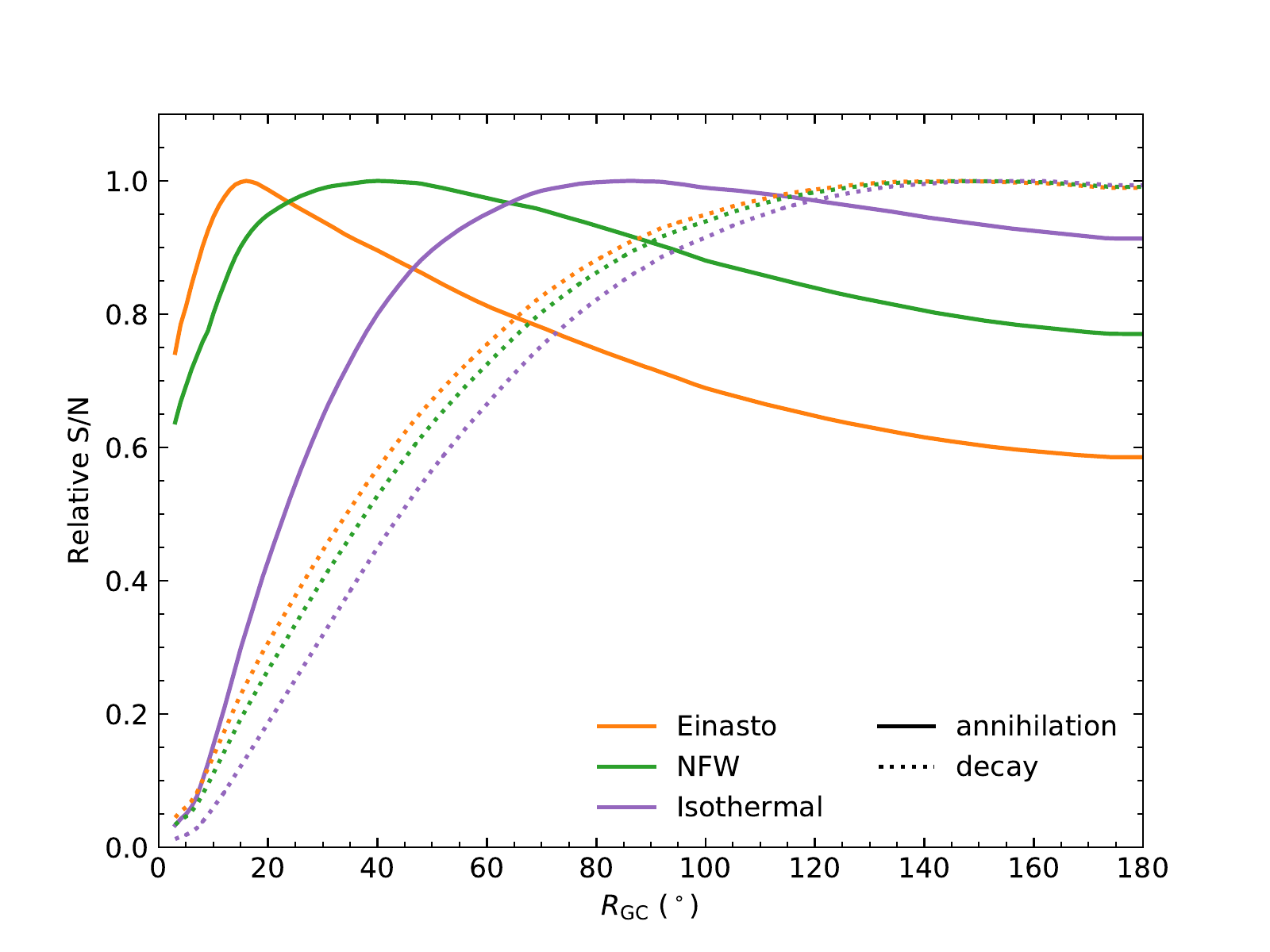}
    \includegraphics[width=0.48\textwidth]{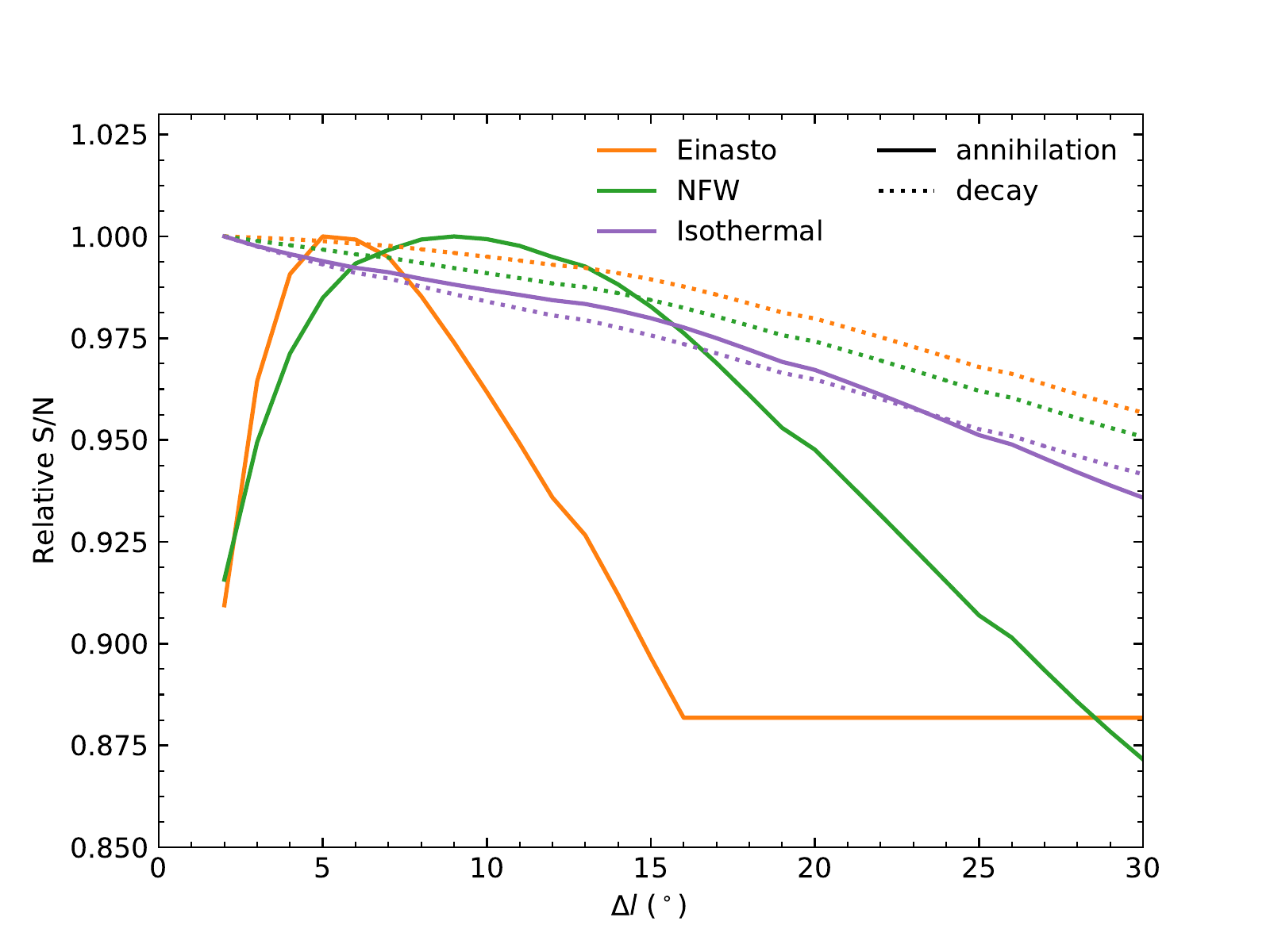}
    \caption{\label{fig:roi_snr}
        The relative S/N values of $R_{\rm GC}$ (left) and $\Delta l$ (right).
        The left panel shows the S/N with respect to the ROI radius $R_{\rm GC}$, with $\Delta l$ fixed to the optimal value for the given DM density profile.
        The solid and dashed lines represent the DM annihilation and decay respectively.
        The right panel shows the ratio as a function of the longitude bound $\Delta l$, with $R_{\rm GC}$ fixed to the optimal value.
    }
\end{figure*}

\begin{figure*}[!htb]
    \centering
    \includegraphics[width=0.7\textwidth]{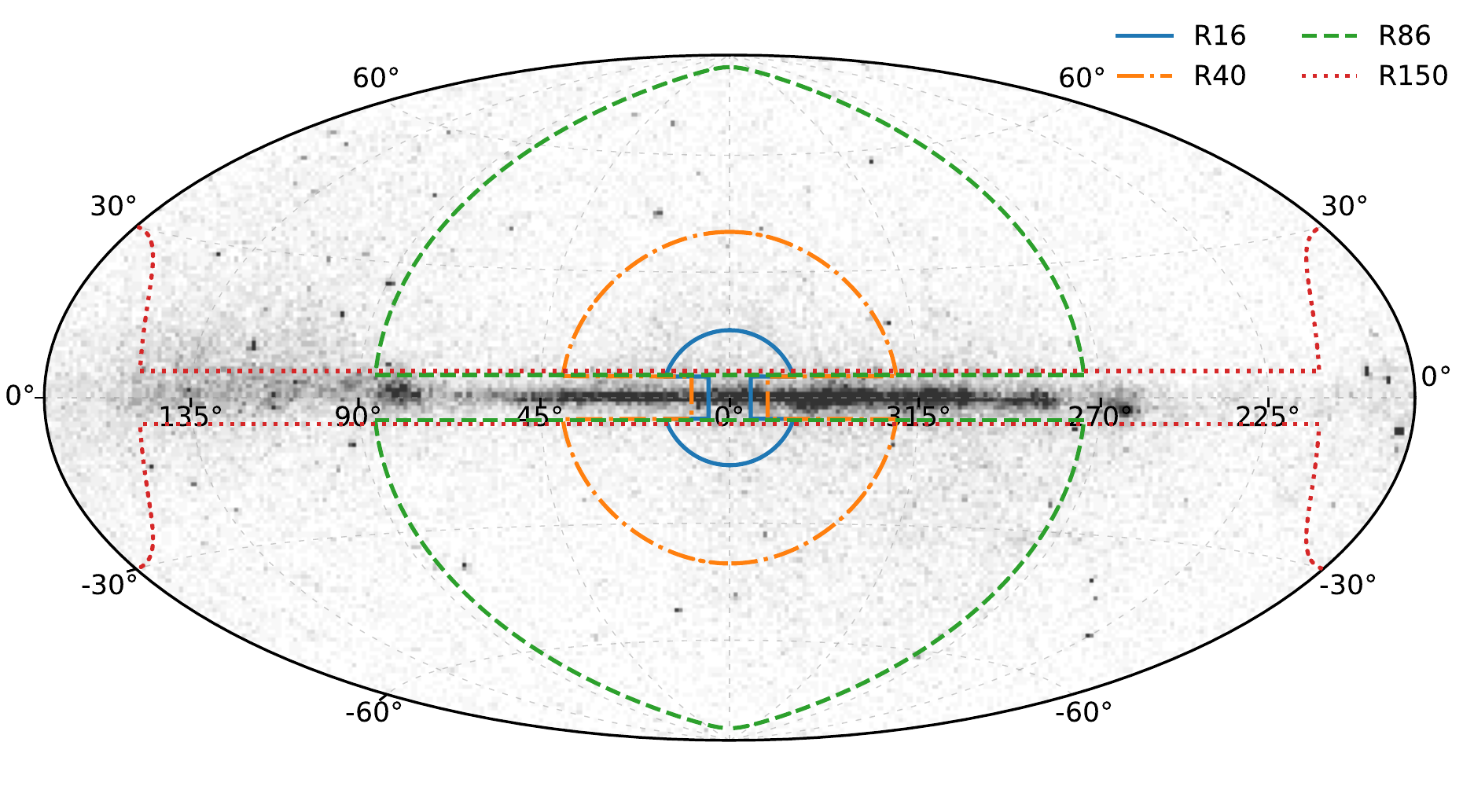}
    \caption{\label{fig:roi}
        The counts map for the LineSearch and BgoOnly data between 5 and 450~GeV binned with the Hammer-Aitoff projection.
        The R16 (blue solid), R40 (orange dot-dashed), and R86 (green dashed) are the optimal ROIs for the Einasto, NFW, and isothermal DM density distribution assuming annihilating DM.
        The R150 (red dotted) is the optimal ROI for the decaying DM with NFW distribution.
    }
\end{figure*}

IRFs describe the performance of the instrument quantitatively and are a necessity to extract the true spectrum and morphology of a source from the observing data.
The IRFs of DAMPE can be divided into three components: the effective area, the point-spread function and the energy dispersion function.
Event-level MC simulations are used to evaluate the IRFs of two types of data sets~\citep{Duan2017}.
Since the effective area and the energy dispersion function are most relevant in our analysis, we introduce them briefly in the following.

The left panel of Fig.~\ref{fig:Aeff_edisp} shows the effective area ($A_{\rm eff}$) at normal incidence, with the solid (dot-dashed) line corresponding to the LineSearch (BgoOnly) data satisfying the HET (High-Energy Trigger) condition.
The $A_{\rm eff}$ of LineSearch data is $\sim 1100~\rm cm^2$ between 10 and 100~GeV.
At the low and high energy ends, the decrease is caused by the lower trigger rate and the backsplash of secondary particles respectively.
The effective area of LineSearch data set at normal incidence is almost the same as that of the baseline one, but is $\lesssim 5\%$ smaller at 30\deg incident angle.
The $A_{\rm eff}$ of BgoOnly data is $\lesssim 360~\rm cm^2$, which is much smaller than that of the STK data because more than half of the BGO converting events do not trigger the detector.
As the trigger rate increases, the effective area peaks at $\sim 100~\rm GeV$.
The effective area is parameterized with a interpolation function of photon energy and incident angle~\citep{Duan2019}.

The right panel of Fig.~\ref{fig:Aeff_edisp} shows the half width of the 68\% energy containment for two data sets at 30\deg incident angle.
The energy resolution above 10~GeV is $\lesssim 1.7\%$ and $\lesssim 1.4\%$ for LineSearch and BgoOnly events respectively.
Both data sets have a resolution better than 1\% above $\sim 35~\rm GeV$.
Since more BGO layers are used in the BgoOnly data set comparing to the LineSearch one in the low energy range, BGO events have a better energy resolution.
The energy dispersion function is parameterized with a piecewise function of incident angle, photon energy and energy deviation~\citep{Duan2019}.

\subsection{Regions of interest (ROIs)}

Signal-to-noise ratio (S/N) optimized ROIs are used for different DM density profiles in order to maximize the sensitivity.
The S/N value is defined to be~\citep{Bringmann2012}
\begin{equation}
    S/N(\mathcal{T}({\bf \Theta})) = \frac{\sum_{i\in \mathcal{T}} \mu_i }{\sqrt{\sum_{i\in\mathcal{T}} c_i}},
    \label{eqn::snr_roi}
\end{equation}
where $\mu_i$ and $c_i$ are the expected and observed photon counts in the $i$-th pixel respectively.
$\mathcal{T}$ represents a pixel set within a parameterized ROI, which is a circular region with a radius of $R_{\rm GC}$ centered on the Galactic center and with a rectangular Galactic plane ($|b|\leq\Delta b$ and $|l|\geq\Delta l$) mask~\citep{Ackermann2013a}.
For the denominator in eqn.~\ref{eqn::snr_roi}, we produce a counts map binned with an order 7 HEALPix scheme~\citep{Healpix2005} using the photons above 5~GeV in the LineSearch data set.
The map is further smoothed with a 2\deg Gaussian kernel to reduce the statistical fluctuation.
We use the product of the $J$-factor ($D$-factor) and the exposure at 5~GeV in the numerator,
which is proportional to the expected photon counts.

We set $\Delta b=5\deg$ to remove the Galactic plane~\citep{Ackermann2013a}, and then scan the remaining parameters ($R_{\rm GC}$ and $\Delta l$) to find the largest ratio.
Fig.~\ref{fig:roi_snr} shows the S/N value with respect to the ROI parameters $R_{\rm GC}$ and $\Delta l$.
For annihilating DM, the cuspy density profiles such as the Einasto and NFW ones tend to require smaller ROI radii (16\deg and 40\deg) in order to reduce the background emission, while the cored isothermal profile prefers a larger radius (86\deg) in order to contain more signal photons.
For decaying DM, a large ROI radius ($\sim 150\deg$) is preferred no matter what density profile it is.
The S/N has much weaker dependence on $\Delta l$.
Einasto and NFW profiles require the optimal $\Delta l$ values of 5\deg and 9\deg for DM annihilation.
For the remaining cases, $\Delta l$ reaches the lower bound (2\deg), so we simply set $\Delta l=0\deg$ to remove all the Galactic plane.

According to the optimal ROI parameters $R_{\rm GC}$ and $\Delta l$, we define the best ROIs for different DM profiles.
The outlines of ROIs are shown in Fig.~\ref{fig:roi} along with the counts map of LineSearch and BgoOnly data set between 5 and 450~GeV.
Our ROIs are very similar to those of \lat except that we do not mask the photons from detected point sources~\citep{LiX2019}, which may slightly weaken our constraints.

\subsection{Systematic uncertainties evaluation with control regions}
The spectral shape of background emission and the inaccurate modeling of exposure are among the major sources of the systematic uncertainties~\citep{Ackermann2015}.
Considering that the photon counts from these systematic uncertainties is usually proportional to the observed one, the uncertainties can be inferred by evaluating the fractional size in control regions~\citep{Ackermann2013a}.
The fractional size $f$, or the so-called ``fractional signal'', is defined to be the ratio of signal counts $n_{\rm sig}$ to the background below the signal peak $b_{\rm eff}$~\citep{Albert2014}.
The ``effective background'' $b_{\rm eff}$ is the square of the statistical uncertainty on the signal counts and can be evaluated with~\citep{Ackermann2015}
\begin{equation}
    b_{{\rm eff},k} = \frac{n_{k}}{\left (\sum_i \frac{F^2_{{\rm sig},ik}}{F_{{\rm bkg},ik}}\right)-1},
\end{equation}
where $n_{k}$ is the total counts in the fit, $F_{{\rm sig},ik}$ and $F_{{\rm bkg},ik}$ are the binned normalized counts distribution functions for signal and background models within the energy bin $i$ of data set $k$ respectively.
The total effective background is then calculated by adding the backgrounds of both data sets together.

In the control regions where no line signal is expected, if a line is detected, it can be caused by the statistical fluctuations and systematic uncertainties.
Therefore the fraction signal can be separated using $f = \delta f_{\rm stat} + \delta f_{\rm sys} = (n_{\rm sig,stat}+ n_{\rm sig,sys})/b_{\rm eff}$.
For the statistical term, $\delta f_{\rm stat}$ can be approximated using the Gaussian distribution $\mathcal{N}(0,1/\sqrt{b_{\rm eff}})$.
Considering the systematic uncertainty is generally proportional to the total counts, the corresponding $\delta f_{\rm sys}$ is a constant.
We can calculate the fractional signal in every energies by performing the same line search analyses in control regions, and then derive the constant part $\delta f_{\rm sys}$.
Once the systematic component $\delta f_{\rm sys}$ is achieved, we can modify the likelihood function to eq.~(5) in the main paper to introduce the systematic uncertainties to the DM constraints.

We choose 5 control regions along the Galactic plane ($|l|>30^\circ$ and $|b|<5^\circ$) where the background emission is similar to that in the ROIs.
In order to balance the statistics, each control region consists of two separate $30^\circ \times 10^\circ$ box-shaped regions with the longitude difference between the box centers being $180^\circ$.
We perform the line search procedures in the control regions and then calculate the effective background and the fraction signal.

\begin{figure}[!htb]
    \centering
    \includegraphics[width=0.48\textwidth]{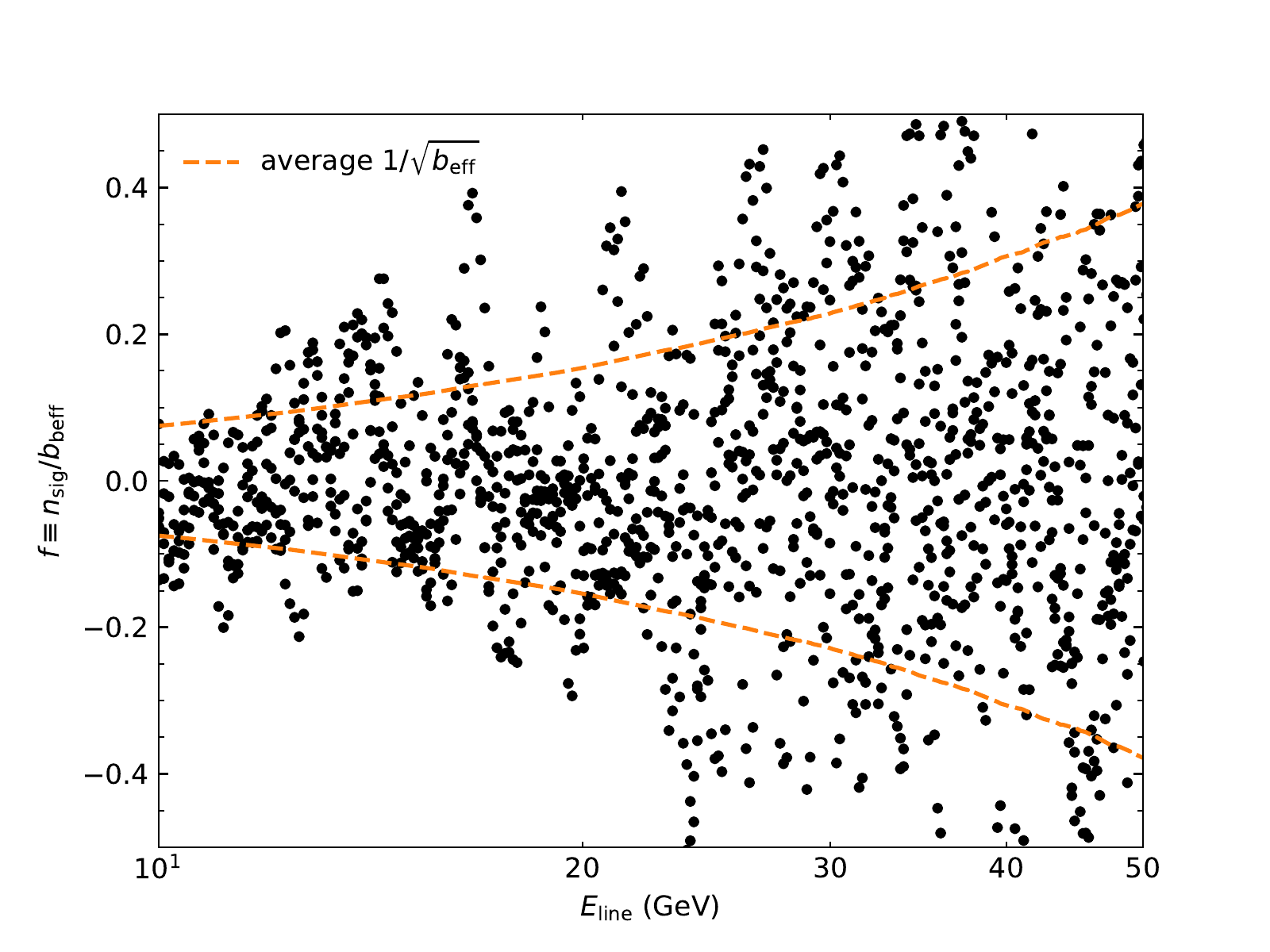}
    \caption{\label{fig:ffac}
        The fractional signal in the control regions.
        Black dots represent the observed $f$ values in 5 control regions along the Galactic plane.
        The orange dashed lines show the average standard deviation of the statistic term $\delta f_{\rm stat}$.
    }
\end{figure}

Fig.~\ref{fig:ffac} presents the fractional signals (black dots) at various line energies from 10 to 50 GeV in all the control regions.
The fractions $f$ mostly scatter within the average statistic uncertainty indicated with the orange dashed lines, which means the fractional signal is dominated by the statistic uncertainties in our energy range.
We acquire the constant systematic term of the fraction by fitting the formula $\chi^2=\sum_i b_{{\rm eff},i} (f_i-\delta f_{\rm sys})^2$.
The optimal $\delta f_{\rm sys}$ is found to be $-0.013$ and $-0.007$ when the fractions below 14~GeV and 20~GeV are adopted respectively.
The negative values mean that the modelled spectrum of Galactic plane is slightly concave in logarithm and it might be caused by the sum of the soft $\pi^0$ decay emission and the hard inverse Compton emission~\citep{Ackermann2012b}.
We adopt the value $|\delta f_{\rm sys}|=0.013$ in the analysis in order to be conservative.

\subsection{The 95\% confidence level limits}
The constraints with the systematic uncertainties addressed are presented in Table~\ref{tab:constraints1} and Table~\ref{tab:constraints2}.

\begin{table}
    \centering
    \caption{\label{tab:constraints1}The 95\% C.L. constraints of DM annihilation cross section or decay lifetime for lines with energies below 100 GeV.
    Please note that the mass of the annihilating DM equals to the line energy, while the mass of the decaying DM is twice the line energy.}
    \begin{tabular}{ccccc}
      \hline
      $E_{\rm line}$ & \multicolumn{3}{c}{$\left < \sigma v \right >_{\gamma\gamma} $ ($\rm cm^3\,s^{-1}$)}
         & $\tau_{\gamma\nu}$ (s) \\
        (GeV) & Einasto & NFW & Isothermal & NFW \\
      \hline
      10.0  &  $9.0 \times 10^{-30}$  &  $2.0 \times 10^{-29}$  &  $7.9 \times 10^{-29}$  &  $1.1 \times 10^{29}$ \\
      10.9  &  $2.0 \times 10^{-29}$  &  $6.3 \times 10^{-29}$  &  $1.0 \times 10^{-28}$  &  $1.7 \times 10^{29}$ \\
      11.7  &  $5.7 \times 10^{-29}$  &  $9.8 \times 10^{-29}$  &  $1.7 \times 10^{-28}$  &  $8.8 \times 10^{28}$ \\
      12.7  &  $2.3 \times 10^{-29}$  &  $3.1 \times 10^{-29}$  &  $1.1 \times 10^{-28}$  &  $2.2 \times 10^{29}$ \\
      13.6  &  $3.7 \times 10^{-29}$  &  $4.6 \times 10^{-29}$  &  $1.4 \times 10^{-28}$  &  $1.5 \times 10^{29}$ \\
      14.7  &  $2.9 \times 10^{-29}$  &  $5.9 \times 10^{-29}$  &  $9.4 \times 10^{-29}$  &  $1.4 \times 10^{29}$ \\
      15.7  &  $2.0 \times 10^{-29}$  &  $4.6 \times 10^{-29}$  &  $1.7 \times 10^{-28}$  &  $1.1 \times 10^{29}$ \\
      16.8  &  $4.9 \times 10^{-29}$  &  $1.1 \times 10^{-28}$  &  $1.7 \times 10^{-28}$  &  $1.5 \times 10^{29}$ \\
      18.0  &  $5.8 \times 10^{-29}$  &  $5.7 \times 10^{-29}$  &  $6.0 \times 10^{-29}$  &  $4.0 \times 10^{29}$ \\
      19.2  &  $9.5 \times 10^{-29}$  &  $9.5 \times 10^{-29}$  &  $5.9 \times 10^{-29}$  &  $4.7 \times 10^{29}$ \\
      20.4  &  $3.3 \times 10^{-29}$  &  $6.7 \times 10^{-29}$  &  $9.1 \times 10^{-29}$  &  $1.6 \times 10^{29}$ \\
      21.7  &  $4.6 \times 10^{-29}$  &  $1.0 \times 10^{-28}$  &  $1.0 \times 10^{-28}$  &  $3.2 \times 10^{29}$ \\
      23.0  &  $4.4 \times 10^{-29}$  &  $8.3 \times 10^{-29}$  &  $1.1 \times 10^{-28}$  &  $2.6 \times 10^{29}$ \\
      24.4  &  $8.9 \times 10^{-29}$  &  $1.4 \times 10^{-28}$  &  $3.8 \times 10^{-28}$  &  $1.4 \times 10^{29}$ \\
      25.8  &  $4.9 \times 10^{-29}$  &  $1.5 \times 10^{-28}$  &  $2.2 \times 10^{-28}$  &  $1.2 \times 10^{29}$ \\
      27.3  &  $4.3 \times 10^{-29}$  &  $6.6 \times 10^{-29}$  &  $2.0 \times 10^{-28}$  &  $1.8 \times 10^{29}$ \\
      28.9  &  $5.8 \times 10^{-29}$  &  $1.6 \times 10^{-28}$  &  $3.0 \times 10^{-28}$  &  $1.0 \times 10^{29}$ \\
      30.5  &  $1.4 \times 10^{-28}$  &  $1.2 \times 10^{-28}$  &  $8.9 \times 10^{-29}$  &  $5.9 \times 10^{29}$ \\
      32.1  &  $5.3 \times 10^{-29}$  &  $6.1 \times 10^{-29}$  &  $2.5 \times 10^{-28}$  &  $1.4 \times 10^{29}$ \\
      33.8  &  $4.3 \times 10^{-29}$  &  $9.3 \times 10^{-29}$  &  $3.3 \times 10^{-28}$  &  $1.1 \times 10^{29}$ \\
      35.6  &  $4.5 \times 10^{-29}$  &  $1.7 \times 10^{-28}$  &  $3.4 \times 10^{-28}$  &  $1.3 \times 10^{29}$ \\
      37.4  &  $2.4 \times 10^{-28}$  &  $2.1 \times 10^{-28}$  &  $4.1 \times 10^{-28}$  &  $1.2 \times 10^{29}$ \\
      39.2  &  $2.2 \times 10^{-28}$  &  $3.6 \times 10^{-28}$  &  $4.4 \times 10^{-28}$  &  $2.1 \times 10^{29}$ \\
      41.1  &  $2.2 \times 10^{-28}$  &  $1.6 \times 10^{-28}$  &  $1.8 \times 10^{-28}$  &  $2.6 \times 10^{29}$ \\
      43.1  &  $6.7 \times 10^{-29}$  &  $2.9 \times 10^{-28}$  &  $3.9 \times 10^{-28}$  &  $1.0 \times 10^{29}$ \\
      45.1  &  $8.5 \times 10^{-29}$  &  $1.7 \times 10^{-28}$  &  $3.8 \times 10^{-28}$  &  $2.9 \times 10^{29}$ \\
      47.2  &  $1.8 \times 10^{-28}$  &  $2.7 \times 10^{-28}$  &  $6.2 \times 10^{-28}$  &  $1.6 \times 10^{29}$ \\
      49.4  &  $2.0 \times 10^{-28}$  &  $3.6 \times 10^{-28}$  &  $1.8 \times 10^{-28}$  &  $4.4 \times 10^{29}$ \\
      51.7  &  $1.6 \times 10^{-28}$  &  $2.8 \times 10^{-28}$  &  $7.2 \times 10^{-28}$  &  $1.4 \times 10^{29}$ \\
      54.0  &  $1.4 \times 10^{-28}$  &  $1.9 \times 10^{-28}$  &  $2.4 \times 10^{-28}$  &  $1.6 \times 10^{29}$ \\
      56.4  &  $1.8 \times 10^{-28}$  &  $1.5 \times 10^{-28}$  &  $3.1 \times 10^{-28}$  &  $3.3 \times 10^{29}$ \\
      58.9  &  $2.4 \times 10^{-28}$  &  $2.3 \times 10^{-28}$  &  $6.7 \times 10^{-28}$  &  $8.5 \times 10^{28}$ \\
      61.4  &  $1.6 \times 10^{-28}$  &  $2.1 \times 10^{-28}$  &  $1.2 \times 10^{-27}$  &  $8.6 \times 10^{28}$ \\
      64.1  &  $3.4 \times 10^{-28}$  &  $6.3 \times 10^{-28}$  &  $5.1 \times 10^{-28}$  &  $1.5 \times 10^{29}$ \\
      66.8  &  $3.0 \times 10^{-28}$  &  $2.3 \times 10^{-28}$  &  $3.7 \times 10^{-28}$  &  $3.6 \times 10^{29}$ \\
      69.6  &  $8.3 \times 10^{-29}$  &  $2.0 \times 10^{-28}$  &  $6.6 \times 10^{-28}$  &  $2.4 \times 10^{29}$ \\
      72.5  &  $9.1 \times 10^{-29}$  &  $3.4 \times 10^{-28}$  &  $3.6 \times 10^{-28}$  &  $1.3 \times 10^{29}$ \\
      75.5  &  $3.9 \times 10^{-28}$  &  $4.1 \times 10^{-28}$  &  $9.8 \times 10^{-28}$  &  $8.2 \times 10^{28}$ \\
      78.5  &  $3.1 \times 10^{-28}$  &  $6.8 \times 10^{-28}$  &  $4.9 \times 10^{-28}$  &  $2.3 \times 10^{29}$ \\
      81.8  &  $5.6 \times 10^{-28}$  &  $6.9 \times 10^{-28}$  &  $8.3 \times 10^{-28}$  &  $1.2 \times 10^{29}$ \\
      85.1  &  $1.9 \times 10^{-28}$  &  $3.5 \times 10^{-28}$  &  $3.2 \times 10^{-28}$  &  $2.9 \times 10^{29}$ \\
      88.6  &  $2.4 \times 10^{-28}$  &  $3.6 \times 10^{-28}$  &  $1.3 \times 10^{-27}$  &  $9.0 \times 10^{28}$ \\
      92.2  &  $8.4 \times 10^{-28}$  &  $6.9 \times 10^{-28}$  &  $6.1 \times 10^{-28}$  &  $2.1 \times 10^{29}$ \\
      96.0  &  $6.7 \times 10^{-28}$  &  $3.7 \times 10^{-28}$  &  $7.7 \times 10^{-28}$  &  $1.4 \times 10^{29}$ \\
      \hline
    \end{tabular}
\end{table}

\begin{table}
    \centering
    \caption{\label{tab:constraints2}The 95\% C.L. constraints of DM annihilation cross section or decay lifetime for lines with energies above 100 GeV.}
    \begin{tabular}{ccccc}
      \hline
      $E_{\rm line}$ & \multicolumn{3}{c}{$\left < \sigma v \right >_{\gamma\gamma} $ ($\rm cm^3\,s^{-1}$)}
      & $\tau_{\gamma\nu}$ (s) \\
     (GeV) & Einasto & NFW & Isothermal & NFW \\
      \hline
      100  &  $2.0 \times 10^{-28}$  &  $5.0 \times 10^{-28}$  &  $7.2 \times 10^{-28}$  &  $2.3 \times 10^{29}$ \\
      104  &  $2.8 \times 10^{-28}$  &  $3.1 \times 10^{-28}$  &  $1.9 \times 10^{-27}$  &  $8.7 \times 10^{28}$ \\
      108  &  $9.4 \times 10^{-28}$  &  $1.1 \times 10^{-27}$  &  $1.3 \times 10^{-27}$  &  $9.2 \times 10^{28}$ \\
      113  &  $2.4 \times 10^{-28}$  &  $6.4 \times 10^{-28}$  &  $1.2 \times 10^{-27}$  &  $8.5 \times 10^{28}$ \\
      118  &  $2.8 \times 10^{-28}$  &  $2.7 \times 10^{-28}$  &  $1.2 \times 10^{-27}$  &  $2.6 \times 10^{29}$ \\
      122  &  $7.4 \times 10^{-28}$  &  $1.0 \times 10^{-27}$  &  $1.2 \times 10^{-27}$  &  $1.8 \times 10^{29}$ \\
      128  &  $5.5 \times 10^{-28}$  &  $3.6 \times 10^{-28}$  &  $6.8 \times 10^{-28}$  &  $3.2 \times 10^{29}$ \\
      133  &  $6.7 \times 10^{-28}$  &  $1.1 \times 10^{-27}$  &  $2.6 \times 10^{-27}$  &  $9.8 \times 10^{28}$ \\
      138  &  $1.1 \times 10^{-27}$  &  $1.3 \times 10^{-27}$  &  $1.0 \times 10^{-27}$  &  $1.1 \times 10^{29}$ \\
      144  &  $9.2 \times 10^{-28}$  &  $1.2 \times 10^{-27}$  &  $1.1 \times 10^{-27}$  &  $2.8 \times 10^{29}$ \\
      150  &  $3.7 \times 10^{-28}$  &  $9.7 \times 10^{-28}$  &  $1.7 \times 10^{-27}$  &  $1.0 \times 10^{29}$ \\
      156  &  $8.9 \times 10^{-28}$  &  $9.7 \times 10^{-28}$  &  $1.9 \times 10^{-27}$  &  $1.1 \times 10^{29}$ \\
      163  &  $4.4 \times 10^{-28}$  &  $5.8 \times 10^{-28}$  &  $4.1 \times 10^{-27}$  &  $9.4 \times 10^{28}$ \\
      169  &  $5.4 \times 10^{-28}$  &  $4.3 \times 10^{-28}$  &  $9.8 \times 10^{-28}$  &  $1.7 \times 10^{29}$ \\
      176  &  $2.8 \times 10^{-27}$  &  $1.5 \times 10^{-27}$  &  $3.5 \times 10^{-27}$  &  $9.1 \times 10^{28}$ \\
      183  &  $7.8 \times 10^{-28}$  &  $2.2 \times 10^{-27}$  &  $2.9 \times 10^{-27}$  &  $1.0 \times 10^{29}$ \\
      191  &  $1.3 \times 10^{-27}$  &  $1.6 \times 10^{-27}$  &  $2.4 \times 10^{-27}$  &  $9.8 \times 10^{28}$ \\
      198  &  $1.4 \times 10^{-27}$  &  $1.1 \times 10^{-27}$  &  $1.1 \times 10^{-27}$  &  $3.1 \times 10^{29}$ \\
      206  &  $9.5 \times 10^{-28}$  &  $1.4 \times 10^{-27}$  &  $2.7 \times 10^{-27}$  &  $7.1 \times 10^{28}$ \\
      214  &  $\cdots$               &  $1.2 \times 10^{-27}$  &  $3.1 \times 10^{-27}$  &  $2.0 \times 10^{29}$ \\
      223  &  $\cdots$               &  $2.6 \times 10^{-27}$  &  $5.4 \times 10^{-27}$  &  $5.5 \times 10^{28}$ \\
      232  &  $\cdots$               &  $3.0 \times 10^{-27}$  &  $3.3 \times 10^{-27}$  &  $8.9 \times 10^{28}$ \\
      241  &  $\cdots$               &  $1.0 \times 10^{-27}$  &  $1.3 \times 10^{-27}$  &  $2.4 \times 10^{29}$ \\
      250  &  $\cdots$               &  $2.6 \times 10^{-27}$  &  $3.7 \times 10^{-27}$  &  $1.1 \times 10^{29}$ \\
      260  &  $\cdots$               &  $2.3 \times 10^{-27}$  &  $3.2 \times 10^{-27}$  &  $2.0 \times 10^{29}$ \\
      270  &  $\cdots$               &  $3.2 \times 10^{-27}$  &  $1.5 \times 10^{-27}$  &  $1.8 \times 10^{29}$ \\
      280  &  $\cdots$               &  $3.9 \times 10^{-27}$  &  $6.8 \times 10^{-27}$  &  $7.2 \times 10^{28}$ \\
      300  &  $\cdots$               &  $3.5 \times 10^{-27}$  &  $3.6 \times 10^{-27}$  &  $1.6 \times 10^{29}$ \\
      \hline
    \end{tabular}
\end{table}


\end{document}